\documentclass[twocolumn]{aastex631}
\usepackage{tabularx}
\usepackage{xspace}
\newcommand{\prot}{\ensuremath{P_{\mbox{\scriptsize rot}}}}
\newcommand{\alphafe}{$[\mathrm{\alpha}/\mathrm{Fe}]$}

\newcommand{\teff}{$T_{\rm eff}$}
\newcommand{\ith}{\ensuremath{^{\rm th}}}

\newcommand{\halpha}{high-$\alpha$}
\newcommand{\logg}{$\log g$\xspace}

\begin{document}

\title{Evidence of Truly Young high-$\alpha$ Dwarf Stars}

\correspondingauthor{Yuxi(Lucy) Lu}
\email{lucylulu12311@gmail.com}

\newcommand{\amnh}{American Museum of Natural History, Central Park West, Manhattan, NY, USA}
\newcommand{\cca}{Center for Computational Astrophysics, Flatiron Institute, 162 5\ith\ Avenue, Manhattan, NY, USA}
\newcommand{\columbia}{Department of Astronomy, Columbia University, 550 West 120\ith\ Street, New York, NY, USA}
\newcommand{\anu}{Research School of Astronomy $\&$ Astrophysics, Australian National University, Canberra ACT 2611, Australia}
\newcommand{\osu}{Department of Astronomy, The Ohio State University, Columbus, 140 W 18th Ave, OH 43210, USA}
\newcommand{\astrothreed}{Center of Excellence for Astrophysics in Three Dimensions (ASTRO-3D), Australia}
\newcommand{\ccapp}{Center for Cosmology and Astroparticle Physics (CCAPP), The Ohio State University, 191 W. Woodruff Ave., Columbus, OH 43210, USA}

\author[0000-0003-4769-3273]{Yuxi(Lucy) Lu}
\affiliation{\osu}
\affiliation{\ccapp}
\affiliation{\amnh}

\author[0000-0001-8196-516X]{Isabel L. Colman}
\affiliation{\amnh}

\author[0000-0001-6180-8482]{Maryum Sayeed}
\affiliation{\columbia}

\author[0000-0003-2508-6093]{Louis Amard}
\affiliation{AIM, CEA, CNRS, Universit\'e Paris-Saclay, Universit\'e de Paris, Sorbonne Paris Cit\'e, Sorbonne Paris Cit\'e, Gif-sur-Yvette, 91191, France}

\author[0000-0002-4031-8553]{Sven Buder}
\affiliation{\anu}
\affiliation{\astrothreed}

\author[0000-0002-0900-6076]{Catherine Manea}
\affiliation{Department of Astronomy, The University of Texas at Austin, 2515 Speedway, Austin, TX, USA}

\author[0000-0002-0842-863X]{Soichiro Hattori}
\affiliation{\columbia}
\affiliation{\amnh}

\author[0000-0002-7549-7766]{Marc H. Pinsonneault}
\affiliation{\osu}
\affiliation{\ccapp}

\author[0000-0003-0872-7098]{Adrian~M.~Price-Whelan}
\affiliation{\cca}

\author[0000-0001-9907-7742]{Megan Bedell}
\affiliation{\cca}

\author[0000-0002-1793-3689]{David Nidever}
\affiliation{Montana State University, P.O. Box 173840, Bozeman, MT, USA}

\author{Jennifer A. Johnson}
\affiliation{\osu}
\affiliation{\ccapp}

\author[0000-0001-5082-6693]{Melissa Ness}
\affiliation{\anu}
\affiliation{\columbia}
\affiliation{\astrothreed}

\author[0000-0003-4540-5661]{Ruth Angus}
\affiliation{\amnh}
\affiliation{\cca}

\author[0000-0002-9879-3904]{Zachary R. Claytor}
\affiliation{Space Telescope Science Institute, 3700 San Martin Drive, Baltimore, MD 21218, USA}
\affiliation{Department of Astronomy, University of Florida, 211 Bryant Space Science Center, Gainesville, FL 32611, USA}

\author[0000-0003-1856-2151]{Danny Horta}
\affiliation{\cca}

\author[0000-0003-0012-9093]{Aida Behmard}
\affiliation{\amnh}
\affiliation{\cca}



\begin{abstract}
The existence of \halpha\ stars with inferred ages $<$ 6 Gyr has been confirmed recently with large spectroscopic and photometric surveys.
However, stellar mergers or binary interactions can induce properties associated with young ages, such as high mass, rapid rotation, or high activity, even in old populations.
Literature studies have confirmed that at least some of these apparently young stars are old merger products.
However, none have ruled out the possibility of genuinely young \halpha\ stars.
Because cool GKM dwarfs spin down, rapid rotation can be used to indicate youth. 
In this paper, we provide strong evidence that truly young \halpha\ stars exist by studying \halpha\ rotators in the Kepler and K2 field with abundance measurements from GALAH and APOGEE. 
After excluding close binaries using radial velocity (RV) measurements from Gaia DR3 and multi-epoch RVs from APOGEE, we find a total of 70 \halpha\ rapid rotators with periods $\sim$10-30 days, 29 of which have lithium measurements from GALAH, indicating that they have not gone through past mass transfer or stellar merger events.
We identify 10 young \halpha\ candidates with no signs of merger-induced mixing or close companions.
One clear example is a G dwarf with a measurable rotation and an age of 1.98$^{+0.12}_{-0.28}$ Gyr that is likely a single star with multiple RV measurements from APOGEE, has significant lithium detection from GALAH (A(Li) = 1.79), and has no signs of planet engulfment.
\end{abstract}



\section{Introduction} \label{sec:intro}
Observation of stars in the Milky Way (MW) disk has shown a bi-modality in the \alphafe--$[\mathrm{Fe}/\mathrm{H}]$ plane that varies across Galactic height and radius \cite[e.g.,][]{Fuhrmann1998, Bensby2011, Nidever2014, Hayden2015}.
The two sequences, namely the high-$\alpha$ and low-$\alpha$ sequences, are also related to the kinematically ``thin'' and ``thick'' disks \cite{Gilmore1983}.
Many theories have been proposed to explain the existence of these two disks \citep[e.g.][]{Chiappini1997, Schonrich2009_b, Clarke2019, Debattista2019, Sharma2020b, Agertz2020, Buck2020, Sahlholdt2022, Lu2022}.
Despite the theory behind the formation of the two sequences, most chemical evolution models and simulations expect stars in the \halpha\ sequence to be old ($>$ 8 Gyr) since it has formed when a large reservoir of gas still exists in the Galaxy and the star formation efficiency was high. 

Interestingly, recent studies on ages of \halpha\ stars show a small number of them ($\sim$ 5\%) younger than 6 Gyr \citep[e.g.,][]{Chiappini2015, Martig2015, SilvaAguirre2018, Claytor2020, Das2020, Zinn2022}.
If these stars are truly young, most of the current chemical evolution models need to be modified to encompass the formation of these young high-$\alpha$ stars (see later in this section for possible formation pathways). 
However, binary interactions such as mergers and mass transfers can easily rejuvenate a star and make it appear young.
As a result, many works in the last decade since the discovery of these young \halpha\ stars have focused on the question: are they products of binary interactions or truly young? 

Some of these apparently young \halpha\ stars are the product of stellar interactions --- they appear to be young because they are more massive and have gained excess mass from their binary companions. 
For example, long-term radial velocity (RV) monitoring of a sub-sample of asteroseismically young \halpha\ APOKASC stars \citep{Pinsonneault2014, Martig2015} suggest at least 50\% of them are in binaries \citep{Jofre2016, Jofre2023}, and some also have indications of debris discs \citep{Yong2016}, agreeing with predictions from theoretical modeling \citep{Izzard2018}.
Moreover, the kinematic studies of these young \halpha\ stars also suggest they are indistinguishable from the rest of the \halpha\ population, indicating a common origin \citep[e.g.,][]{Zhang2021, Jofre2023}. 
\cite{Miglio2021} find that the number of young \halpha\ (equivalent to more massive) stars occurs more frequently among red clump stars compared to red giant branch stars.
Since red clump star is a later evolutionary stage than red giant branch star, if young \halpha\ stars are formed from interactions with companion stars, the occurrence rate of such stars would be higher in the red clump phase since they would have more time to undergo mass-transfer.
Many of these red giant branch stars also exhibit abnormal chemical abundances such as $[\mathrm{C}/\mathrm{N}]$ and $[\mathrm{Ba}/\mathrm{Fe}]$ ratios \citep[e.g.,][]{Hekker2019, Zhang2021, Jofre2023} and faster rotational velocity \citep[e.g.,][]{Tayar2015}.
Very recently, \cite{Yu2024} also discovered that a higher fraction of young \halpha\ subgiant stars exhibit heightened stellar activity, hinting they have experienced binary interactions.
Despite the clear evidence that some young \halpha\ stars are products of binary interactions, not all exhibit distinguishable chemical signatures from the rest of the \halpha\ population \citep[e.g.,][]{Matsuno2018}, suggesting if mass transfer is the sole explanation for these young \halpha\ stars, at least a subset of them either went through mergers on the main-sequence or have only gained mass from a low-mass red giant or main-sequence star companion. 

However, in detail, problems with the assumption of a purely old population emerge.  
First, the fraction of mergers is substantially greater than what is predicted by binary synthesis models \citep{Izzard2018}.  
Second, as noted in \cite{Jofre2016}, merger models predict high [C/N] in general.  
This is because the progenitors were lower-mass stars with cooler cores; a late main-sequence or post-main-sequence transfer event will have a smaller CN-processed core than naturally massive stars.
Many of the \halpha\ stars do indeed have high [C/N], but many do not \citep{Jofre2023}. 
Interestingly, as noted in that paper, virtually all young \halpha\ giants are C-rich, which could indicate a distinct nucleosynthetic origin. 
If we treat the \halpha\ population as a mixed one, the population arguments above weaken; the presence of \textit{some} old stars in the group does not require \textit{all} stars to be old.

To this date, none of the evidence mentioned above can exclude the existence of genuinely young \halpha\ stars. 
Recently, \cite{Sun2023} discovered a group of O-rich subgiant stars between 2-4 Gyr using GALAH \citep{DeSilva2015} that they attribute to the increase in star formation induced by the infalling of the Sagittarius dwarf galaxy \citep{Ibata1994}.
However, this group of young O-rich stars seems to exhibit similar $[\mathrm{Mg}/\mathrm{Fe}]$ values as the low-$\alpha$ disk, making it difficult to conclude whether this group of stars can be associated with the young \halpha\ stars.
\cite{Horta2024} reported 7 halo stars between 6-9 Gyr old that are associated with the Gaia-Sausage-Enceladus \citep[GSE;][]{Belokurov2018, Helmi2018} merger using the subgiant age catalog from \cite{Xiang2022} that are possibly younger than most stars in the GSE (8-11 Gyr).
If intermediate-age halo stars exist, young \halpha\ stars can potentially form in similar processes. 
Some possible formation scenario includes clumpy formation \citep[e.g.,][]{Garver2023}, complex chemical evolution near the bar's co-rotation resonance \citep{Chiappini2015}, large fluctuations in the Type Ia Supernova rate at large Galactic Radius \citep{Johnson2021}, or recent accretion event \citep{Sun2023}.

Most detailed studies on young \halpha\ stars have been focusing on evolved stars, and only a handful have been done with main-sequence stars. 
This is most likely due to the difficulties in age-dating main-sequence stars, as their temperature and luminosity change extremely slowly over time.  
Luckily, gyrochronology \citep{Barnes2003} has been shown to reliably age-date main-sequence stars using rotation period measurements \citep[e.g.,][]{Angus2015, Matt2015, vansaders2016, Garraffo2018, Spada2020, Curtis2020, Claytor2020, Hall2021, Lu2021, SilvaBeyer2022, Pass2022, Bouma2023, Lu2024}. 
If all \halpha\ stars are of age 7 Gyr and older, the \halpha\ dwarf stars would be rotating on the order of at least 30 days or even slower based on the empirical gyrochronology model presented in \cite{Lu2024}. 
However, \cite{Claytor2020} has found a small group of fast rotating (hence young according to gyrochronology) \halpha\ stars using the Kepler \citep[][]{Borucki2010} and mentioned that since interactions between close binary companions can spin-up stars via tidal interactions, these young \halpha\ dwarf stars might not be genuinely young.
Thus, the same question to the evolved stars exists --- are these fast-rotating \halpha\ stars created from binary interactions, or are they truly young? 
If these stars have spun up by close-by binary companions \citep[typically binaries in $<\sim$ 15-day orbital period][]{Lurie2017}, gyrochronology can predict ages that are 200\% younger than their true age \citep[e.g.,][]{Fleming2019}. 
As a result, accounting for binary interactions is also important in searching for intermediate-age \halpha\ dwarf stars. 

One other concern regarding binary interactions is stellar mergers --- a star that appears to be single can also be the result of a stellar merger event.
In this case, the surface of the star can also be spun up during such event.
Fortunately, unlike evolved stars, lithium depletion is a strong indicator for \textit{stellar} merger or mass transfer for stars on the main-sequence \citep[e.g.,][]{Ryan2001, Pinsonneault2002, Ryan2002}, and it has long been used as evidence that blue stragglers are formed from binary interactions \citep[e.g.,][]{Pritchet1991, Glaspey1994}.
Lithium is destroyed at a relatively low temperature of 2.5 $\times$ 10$^6$ K.
It survives only in the surface layer of main-sequence or turn-off stars, in which the thin layer only makes up a few percent of the stellar mass.
Thus, radial mixing induced by binary mergers or mass transfer can cause significant lithium depletion on the stellar surface. 
As a result, if we can find fast-rotating main-sequence single stars with normal lithium abundances, these stars are either truly young or have only engulfed something of hot Jupiter mass, which would not induce enough mixing to destroy lithium.
\citep{Poppenhaeger2014} shows a hot Jupiter planet engulfment event can spin up a star by up to 15 days.
However, hot jupiters are likely rare and the planet occurrence rate also decreases with decreasing metallicity \citep{Fischer2005}.
As a result, young \halpha\ stars with metallicity $<$ 0 dex likely did not get spun up by engulfing a hot jupiter.

In this paper, we present the strongest evidence thus far that at least a fraction of young \halpha\ stars is likely genuinely young by studying rotating dwarf stars in the Kepler field \citep{Borucki2010} and K2 \citep{Howell2014} fields with spectroscopy from The Apache Point Observatory Galactic Evolution Experiment \citep[APOGEE;][]{Majewski2017} and GALAH.
After excluding close binaries, we find 41 APOGEE single stars with rotation periods measured from Kepler and K2 and 29 GALAH stars with period measurements from K2 that also have lithium abundances reported by \cite{Wang2024} (see Section~\ref{sec:dataselection}).
All these stars have rotation $<$ 40 days, indicating their youth.
The age, kinematic, and abundance distributions for these stars are discussed in Section~\ref{subsec:agedist}, Section~\ref{subsec:kinematic}, and Section~\ref{subsec:abundance}, respectively.
In Section~\ref{subsec:period}, we used the \prot-\teff\ diagram and selected 10 prime candidates for truly young \halpha\ stars, in which we performed a deeper analysis on one G dwarf that has both APOGEE and GALAH observations.
In detail, this G dwarf has an age of 1.98$^{+0.12}_{-0.28}$ Gyr, inferred from \texttt{STAREVOL} \citep{Siess2000, Amard2019}, and is likely a single star based on multiple APOGEE RV measurements, consistent stellar parameters inferred from APOGEE and GALAH, no signs of planet engulfment, and have significant and excess lithium measurement from GALAH compared to stars of similar \teff, \logg, $[\mathrm{Fe}/\mathrm{H}]$, and $[\mathrm{Mg}/\mathrm{Fe}]$, hinting a young nature for this G dwarf (see Section~\ref{subsec:trulyyoung}).
Finally, A discussion on the treatment of binaries can be found in Section~\ref{sec:binaires}.

\section{Data \& Selection} \label{sec:dataselection}
To search for truly young \halpha\ dwarf stars, we are ideally in search of stars 1) that are rotating with a period less than $\sim$30 days, 2) with no close binary companions, and 3) with surface lithium measurements.
For 1), based on spin-down modelings, all \halpha\ dwarf stars should be rotating so slowly that they would either be detected as a very slow rotator \citep[for reference, a 7 Gyr star would be rotating around 30 days using the model presented in][]{Lu2024} or not at all. 
As a result, if a \halpha\ star is rotating for less than 30 days, it is potentially intermediate-aged or young.
For 2), a fast-rotating star could have gotten spun up by a close-in binary companion.
Studies show that there is a population of old synchronized binaries with periods $<$ 10-15 days \citep{Meibom2005, Lurie2017, Simonian2019} are in synchronized binary systems. 
As a result, it is important to take into account the effects of binaries.
For 3), even if a star is single, it could still have gone through a stellar merger that has also spun up the star. 
However, this scenario can be excluded if it still has significant surface lithium measurements as lithium is easily destroyed during stellar merger events.

To search for fast-rotating (young) \halpha\ dwarf stars, we started with two samples --- 1) The APOGEE--K2/Kepler sample: \halpha\ Kepler and K2 stars with APOGEE DR17 \citep{APOGEEDR17} reduced, wavelength-calibrated, co-added spectra \citep{APOGEE_pipeline}, and abundance measurements using the reduced spectra with the APOGEE Stellar Parameters and Chemical Abundances Pipeline \citep[ASPCAP;][]{ASPCAP_1, ASPCAP_2}, and 2) The GALAH--K2 sample: \halpha\ K2 stars with GALAH DR3 spectra \citep{Galahdr3} and chemical abundances derived through Spectroscopy Made Easy \citep[SME;][]{Valenti1996, Piskunov2017}.
Ideally, we aim to find single, rotating dwarf stars with periods $<$ 30 days and with lithium measurements. 
A confirmed lithium measurement for a dwarf star can exclude past stellar mass-transfer or stellar merger events, and if this star is also rotating relatively fast and is not in a close-in binary system, the most possible scenarios are either it is truly young, or that it has engulfed a planet. 
More discussion on how binaries can affect our results is included in Section~\ref{sec:binaires}.

\subsection{Selection of \halpha\ dwarf stars}\label{subsec:spectra}
We first did an initial exclusion of binaries with Gaia DR3 \citep{Gaia, Gaiadr3} by excluding any stars with the Gaia Renormalized Unit Weight Error (RUWE) $<$ 1.2.
However, it is worth pointing out that the binary threshold can vary across the sky \citep{CastroGinard2024} and the HR diagram \citep{Penoyre2022}.
Given the type of stars (main-sequence GK dwarfs) and their sky positions, the RUWE value threshold of 1.2 should be sufficient to perform the initial binary cut. 
We also excluded stars with \texttt{ipd\_frac\_odd\_win} and \texttt{ipd\_frac\_multi\_peak}\footnote{description of these parameters, see \url{https://gea.esac.esa.int/archive/documentation/GDR3/Gaia_archive/chap_datamodel/sec_dm_main_source_catalogue/ssec_dm_gaia_source.html}} $>$ 0, as they could indicate unresolved binaries \citep{Lindegren2021}.
We then performed additional quality cuts based on the descriptions below for APOGEE (Section~\ref{subsubsec:apogee}) and GALAH (Section~\ref{subsubsec:galah}). 
Lastly, since hot stars do not generate spots or have measurable lithium, we also constrain our sample to stars with \teff\ $<$ 6,500 K.

\subsubsection{For APOGEE}\label{subsubsec:apogee}
\begin{figure*}
    \centering
    \includegraphics[width=\linewidth]{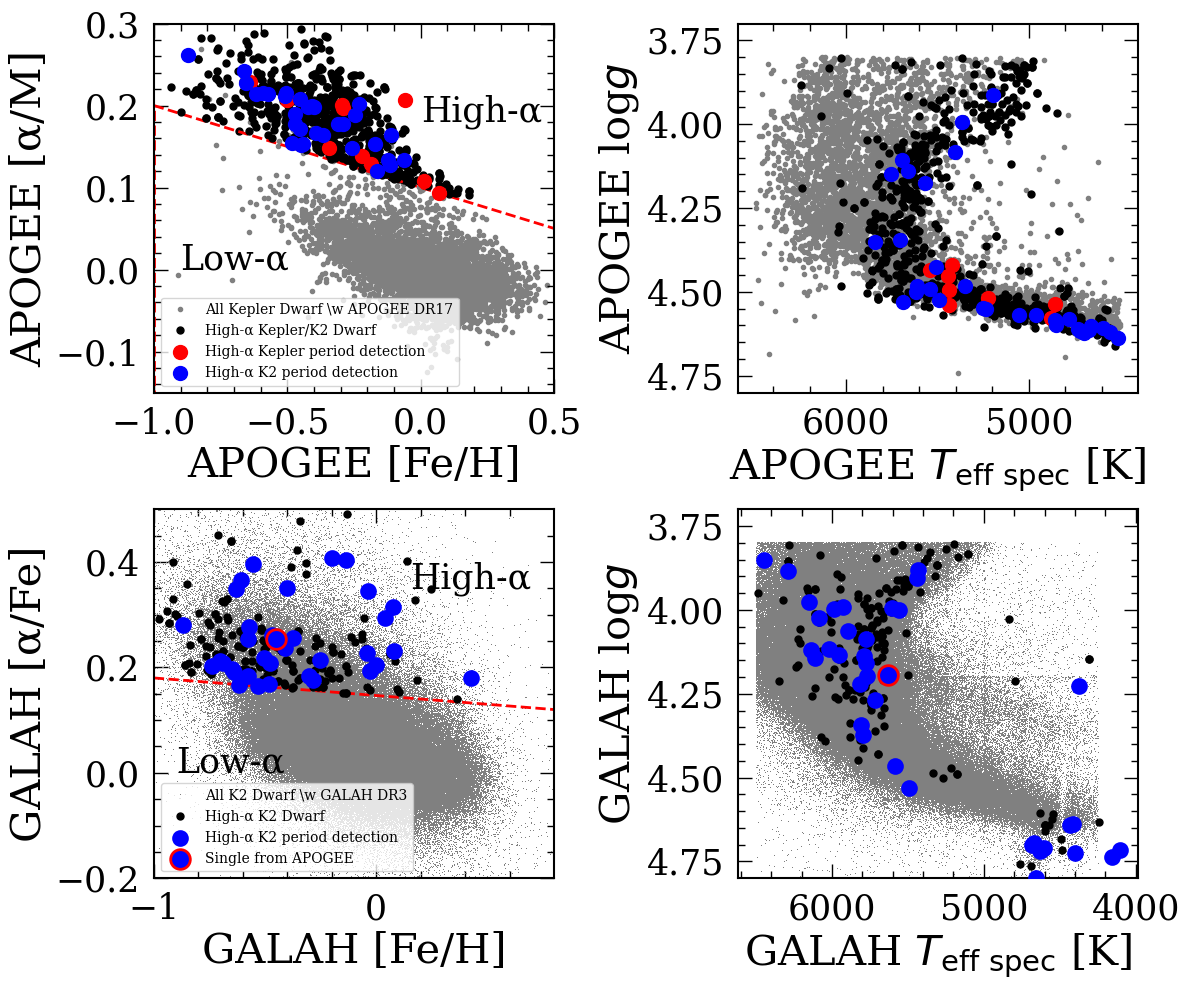}
    \caption{The APOGEE--K2/Kepler sample (top row) and the GALAH--K2 sample (bottom row) shown in the $[\mathrm{Fe}/\mathrm{H}]$-\alphafe\ plane (left column) and \logg-\teff\ plane (right column). 
    The dashed red lines in the $[\mathrm{Fe}/\mathrm{H}]$-\alphafe\ planes show the division between low- and \halpha\ disk for APOGEE and GALAH.
    The two lines are defined separately. 
    The black points in the top plots show the full 777 APOGEE--K2/Kepler stars that we searched for rotation signal, and the red and blue points show the 41 stars with detected periods in the APOGEE--K2/Kepler sample. The black points in the bottom plots show the full 232 GALAH--K2 sample, and the blue points show the 37 stars with period detection from the GALAH--K2 sample.
    The point with a red outline shows the star with both APOGEE and GALAH spectra that we performed a detailed analysis of (see Section~\ref{subsec:trulyyoung}).
    The background grey points show the full cross-match of APOGEE--K2/Kepler and GALAH--K2 including all the low-$\alpha$ stars.
    It is worth pointing out that we excluded binaries with individual APOGEE RV measurements for the APOGEE sample (top row), and we only selected stars with lithium measurements from \cite{Wang2024} for the GALAH sample (bottom row).}
    \label{fig:fig1}
\end{figure*}

The APOGEE observing strategy provides multiple RV measurements for a large fraction of stars.
As a result, for the APOGEE--K2/Kepler sample, we selected stars with more than two visits in the hope of further excluding stars that are in close-in binary systems. 
The quality cuts on the APOGEE DR17 parameters are:
\begin{itemize}
    \item 4,500 K $<$ Effective Temperature (\teff) $<$ 6,500 K, to select stars with reliable abundances and the possibility of measuring periods. 
    \item Signal-to-noise (SNR) $>$ 50, to select stars with high-quality spectra.
    \item \logg\ $>$ 3.8, to select dwarf stars. 
    \item \alphafe\ and $[\mathrm{Fe}/\mathrm{H}]$ ASPCAP flags = 0, to select reliable $[\mathrm{Fe}/\mathrm{H}]$ and \alphafe\ measurements. 
    \item Number of visits $>$ 2, RV scatter between the visits $<$ 1 km/s, to select stars likely without close-in binary companions. 
\end{itemize}
To select \halpha\ stars, we performed a cut with a line defined by ($[\mathrm{Fe}/\mathrm{H}]$, \alphafe) = (-0.5, 0.5) and  ($[\mathrm{Fe}/\mathrm{H}]$, \alphafe) = (0.15, 0.05). 
This line is indicated with the red dashed line in Figure~\ref{fig:fig1} top left plot. 
We then cross-matched the APOGEE \halpha\ star sample with the Kepler and K2 stars with a 1.2'' cone search, which yielded 317 and 460 APOGEE stars that are in Kepler and K2, respectively (shown as black points in Figure~\ref{fig:fig1} top plots).

\subsubsection{For GALAH}\label{subsubsec:galah}
With high-resolution optical spectra, GALAH is able to observe the 6707.814 \AA\ lithium line \citep{Smith1998}.
Recently, \cite{Wang2024} measured lithium abundances for $\sim$ 600,000 stars in GALAH DR3 \citep{Galahdr3} using non-local thermodynamic equilibrium (NLTE) radiative transfer that considers 3D effects in the model atmospheres.
Since we are in search of single dwarf stars with lithium measurements, we selected stars applying the following quality cuts:
\begin{itemize}
    \item 4,000 K $<$ \teff\ $<$ 6,500 K, to select stars with reliable abundances and the possibility of measuring periods.
    \item SNR $>$ 50, to select stars with reliable spectra. 
    \item log$g$ $>$ 3.8, to select dwarf stars.
    \item With lithium abundance measurements reported from \cite{Wang2024}.
\end{itemize}

Unfortunately, the stars in our selected sample have only been visited once, as a result, some of the GALAH stars are likely in unresolved binaries. 
Later on, for the GALAH--K2 stars with period measurements, we also excluded binaries for stars with multiple APOGEE visits and stars that are double-lined spectroscopic binaries. 

Since abundance measurements from various surveys can exhibit systematic offsets and disagreements \citep[e.g.,][]{Hegedus2023}, we defined the line that distinguishes the high- and low-$\alpha$ disks separately for the GALAH sample.
We defined this line with ($[\mathrm{Fe}/\mathrm{H}]$, \alphafe) = (-1, 0.5) and ($[\mathrm{Fe}/\mathrm{H}]$, \alphafe) = (0.18, 0.13), as indicated with the red dashed line in Figure~\ref{fig:fig1} bottom left plot. 
We then performed the same 1.2'' cone search cross-match with K2, which provided us with 232 GALAH--K2 stars with lithium measurements from \cite{Wang2024}. 
This sample is shown as black points in Figure~\ref{fig:fig1} bottom plots.

In the right panels of Figure~\ref{fig:fig1}, the bulk of the \halpha\ population traces out the right edge of the field Kiel diagram; this is where we would expect a uniformly old population to be observed. 
The small number of points to the left are the candidate merger / young population objects. 
Note that mergers can also be found redwards of the field turnoff.  
In the GALAH sample on the lower right, we can already see some interesting results.  
The stars in the upper left of the diagram do not appear to have significantly lower lithium than those closer to the field turnoff; if all were mergers or have experienced mass transfer, By contrast, for halo stars, essentially all stars hotter than the field turnoff are severely depleted in lithium relative to the Spite plateau \citep[e.g.,]{Ryan2001, Ryan2002}.

\subsection{Rotation Period measurements}
\subsubsection{For Kepler Stars}

We searched the sample of 317 APOGEE--Kepler \halpha\ targets with 30-minute cadence Kepler data, using the simple aperture photometry (SAP) data and performing corrections using the Python-based software pipeline described in \citet{Colman17}. We prepared each quarter individually, removing systematic trends by high-pass filtering, dividing by a smoothed light curve produced by convolving the flux with a Gaussian kernel with a width of 250 cadences, or $\sim$5 days, and then removing outlying points beyond 3$\sigma$ of the mean flux. We analyzed the resulting light curves using a Lomb-Scargle periodogram as implemented in \texttt{astropy} \citep{astropy:2013, astropy:2018, astropy2022}. We then automatically phase-folded each light curve on the peak period at maximum power.

Due to the small number of targets, we could feasibly perform visual inspections of each stitched light curve. For stars that appeared to have longer periods, we performed a follow-up inspection of all candidate detections, producing alternate light curves filtered with 100-cadence and 500-cadence Gaussian kernels to confirm the absence of systematic. In the few cases where the peak at maximum power was not the clearest period detection in the periodogram --- usually due to low-frequency, high-amplitude noise --- we were able to then manually fold on a selected period and confirm its significance.
We recovered periods for $\sim$ 15 stars from the APOGEE--Kepler sample, equivalent to a yield of $\sim$ 5\%. 

Within our full sample of stars that we searched for periods with Kepler, 10 has reported periods from \cite{McQuillan2014} and \cite{Santos2021}, ranging from 15-50 days.
This suggests our period recovery rate is similar to that of literature for the \halpha\ disk.
The recovery rate is low compared to the full Kepler sample because most Kepler field stars are intermediate-age low-$\alpha$ stars, and we selected stars that are in the \halpha\ disk, which if are all old, should not have easily detectable rotation periods. 
Compared to the literature rotation period measurements, our period search was able to recovered the periods of 6 stars out of the 10.
We measure the periods of these 6 stars to be exclusively half-periods of the reported periods.
However, this is not surprising as the stars with longer periods ($>$ 40 days) will have low amplitudes and thus, harder to confirm by eye.

\subsubsection{For K2 Stars}
To obtain rotation periods from K2, we queried the default K2 light curve using \texttt{lightkurve} \citep{lightkurve2018}.
We used the Pre-search Data Conditioned Simple Aperture Photometry (PDCSAP) flux as recommended \citep{Howell2014}. 
To get rid of systematic offsets, we first divided the flux by the median flux value and subtracted one so that the flux measurements were centered around 0.
We then subtracted a linear fit to the normalized flux. 
To obtain the periods, we used the one-term Lomb--Scargle periodogram function from \texttt{astropy}, with a \texttt{samples\_per\_peak} (desired samples across the typical peak) of 100. 

\begin{figure*}
    \centering
    \includegraphics[width=\textwidth]{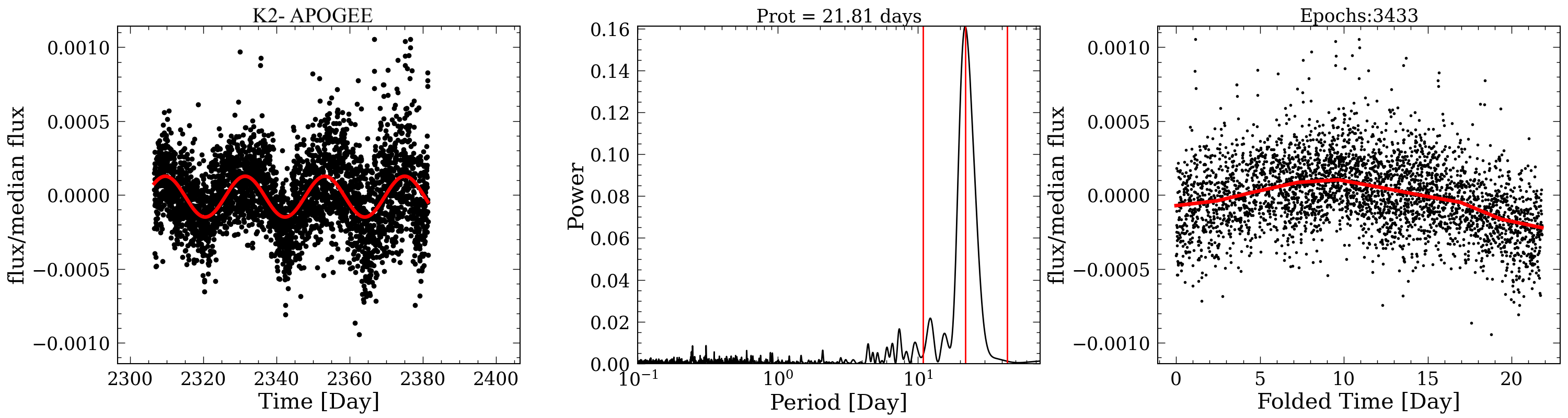}
    \includegraphics[width=\textwidth]{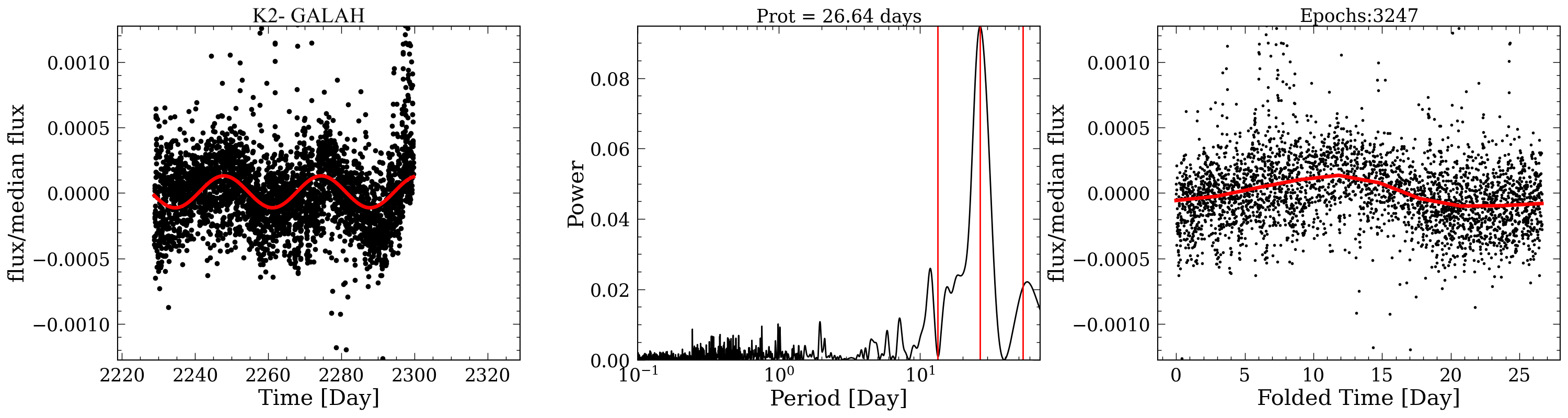}
    \caption{Period detection for an APOGEE--K2 (top) and a GALAH--K2 example star (bottom) that does not have a maximum Lomb--Scargle power $>$ 0.1. 
    Left column: Lomb--Scargle model fits to the detected periods (red lines) over-plotting on the K2 PDCSAP light curves.
    Middle column: Lomb--Scargle periodograms of the light curves on the left with the detected periods and the half-period harmonics marked by the red vertical lines. 
    Right columns: Folded light curves onto the detected period with a sliding window median plotted in red on top.
    The bottom plots show the period detection for one of the possibly genuinely young \halpha\ stars.
    }
    \label{fig:K2_lk}
\end{figure*}

We vetted all K2 light curves by eye and required a typical maximum peak power $>$ 0.1 unless a very clear signal is spotted.
Figure~\ref{fig:K2_lk} shows two examples where the peak powers are $<$ 0.1 but are identified as rotators for an APOGEE--K2 (top) star and a GALAH--K2 star (bottom). 
The GALAH--K2 star shown in the bottom plot is also the star (Gaia DR3 47490179242744960) identified later that is our strongest candidate for being a genuinely young star (see Section~\ref{subsec:twostars} for detail).
We recovered periods for $\sim$ 8\% of the K2 stars, compared to $\sim$13\% for the full K2 sample \citep{Gordon2021}.

\subsection{\halpha\ GALAH--K2 and APOGEE--K2/Kepler stars with period measurements}\label{subsec:GALAHAPOGEE}

\begin{figure}[h]
    \centering
    \includegraphics[width=\columnwidth]{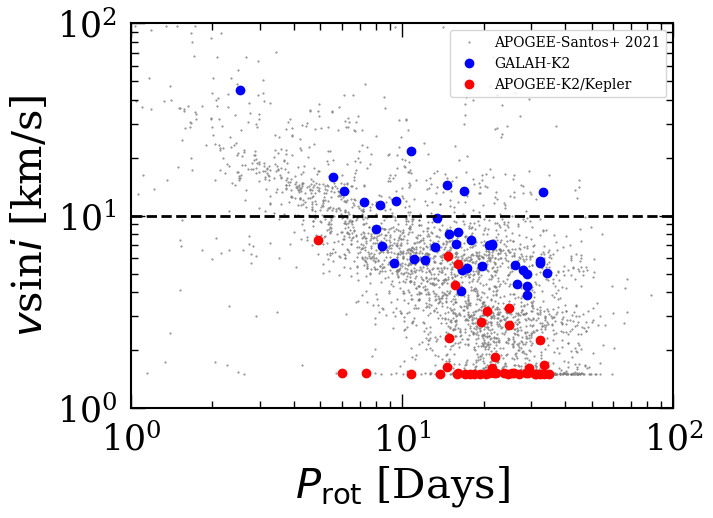}
    \caption{Period measurements in this work plotted against the reported \texttt{VSINI} values for stars in APOGEE and \texttt{vbroad} values for stars in GALAH.
    The overlapping sample of APOGEE and period measurements from \cite{Santos2021} and \cite{McQuillan2014} are shown as the background grey points.
    The dashed black line shows the APOGEE $v\sin{i}$ detection limit of 10 km/s.
    Overall, our distribution matches well with that of the literature.}
    \label{fig:vsini}
\end{figure}

Finally, for stars with period measurements, we excluded those that have any identified Gaia sources within 5'' to account for possible contamination.
We also excluded any sources that are identified as variable stars or binaries according to Simbad \citep{simbad}.
This left us with 41 APOGEE--K2/Kepler stars (10 from Kepler and 31 from K2) and 37 GALAH--K2 stars with period measurements.
\prot-$v\sin{i}$ measurements of these stars are shown in Figure~\ref{fig:vsini}, and their locations in \alphafe-$[\mathrm{Fe}/\mathrm{H}]$ and \logg-\teff\ are shown in Figure~\ref{fig:fig1}. 

Within our sample, 15 K2 stars have both APOGEE and GALAH spectra available.
For stars with lithium measurements from GALAH, 3 have visited more than once by APOGEE and have RV scatter $>$ 1 km/s, and thus are excluded from our final sample.
Besides these 3 stars, we also excluded another 3 stars from the full GALAH--K2 rotator sample as they are identified as double-lined spectroscopic binaries according to \cite{Traven2020}. 

\begin{figure}[h]
    \centering
    \includegraphics[width=\columnwidth]{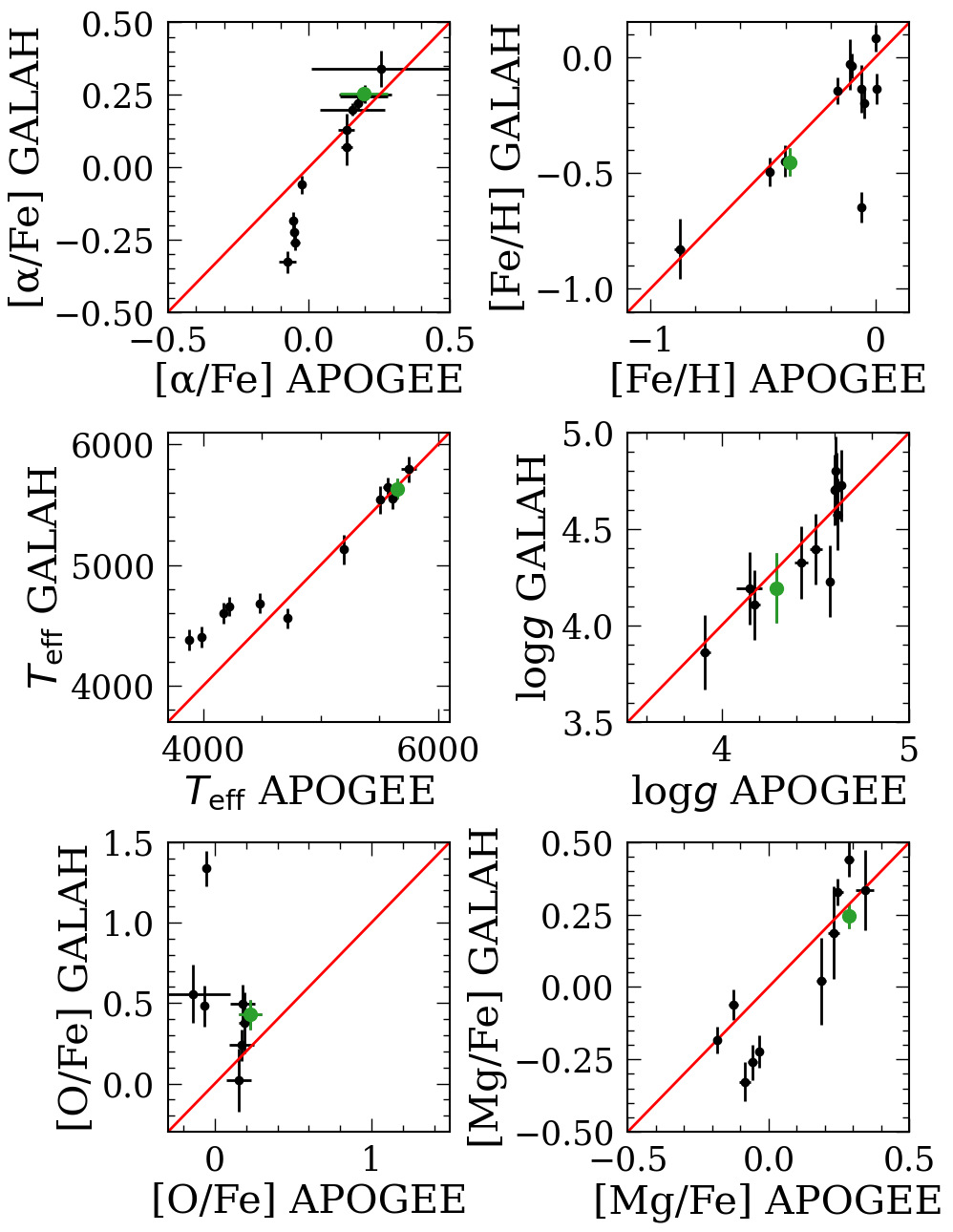}
    \caption{Comparison of the stellar parameters between the 12 rotators in our sample that have both GALAH DR3 and APOGEE DR17 spectra and abundances. 
    The green point shows the star with significant lithium measurement in GALAH DR3 spectra and is likely a single star based on APOGEE multiple RV measurements (see Section~\ref{subsec:trulyyoung}).
    This star also has abundances agreeing between the two surveys. 
    The uncertainty for \alphafe\ measurements from APOGEE is calculated by adding the uncertainty from $[\mathrm{\alpha}/\mathrm{M}]$, $[\mathrm{Fe}/\mathrm{H}]$, and $[\mathrm{M}/\mathrm{H}]$ in quadrature.
    In general, \teff, \logg, \alphafe, $[\mathrm{Fe}/\mathrm{H}]$, and $[\mathrm{Mg}/\mathrm{Fe}]$ agree within the reported uncertainty with a slight trend at low \teff, $[\mathrm{Mg}/\mathrm{Fe}]$, and \alphafe.
    $[\mathrm{O}/\mathrm{Fe}]$ measurements from GALAH and APOGEE do not agree, which is expected.}
    \label{fig:fig_galah_apogee}
\end{figure}

Excluding the three GALAH--K2 binaries without stellar parameter measurements from APOGEE DR17, Figure~\ref{fig:fig_galah_apogee} shows the comparisons between stellar parameters for the 12 remaining stars that are in both APOGEE and GALAH. 
In general, \teff, \logg, \alphafe, $[\mathrm{Fe}/\mathrm{H}]$, and $[\mathrm{Mg}/\mathrm{Fe}]$ agree within the reported uncertainty between APOGEE and GALAH.
However, there is a slight trend towards lower values of \alphafe, \teff, and $[\mathrm{Mg}/\mathrm{Fe}]$. 
However, $[\mathrm{O}/\mathrm{Fe}]$ measurements from GALAH and APOGEE do not agree, which is expected \citep[see also Figure 3 from][]{Griffith2019} as oxygen is hard to measure in optical spectra since the O triplet lines are strongly affected by 3D non-local thermodynamic equilibrium effects \citep[e.g.,][]{Kiselman1993, Amarsi2016}.
It is also worth noting that some \halpha\ stars defined using the ASPCAP abundances are considered low-$\alpha$ stars using GALAH abundances (see the top left plot in Figure~\ref{fig:fig_galah_apogee}), further showing the difficulties in obtaining element abundances from stellar spectra as well as defining stellar populations combining different spectroscopy surveys. 
Nonetheless, most stellar parameters agree within uncertainty between GALAH and APOGEE. 

\subsection{GALAH--K2 rotators with Gaia DR3 RVs}\label{subsec:galahgaia}
\begin{figure*}
    \centering
    \includegraphics[width=\textwidth]{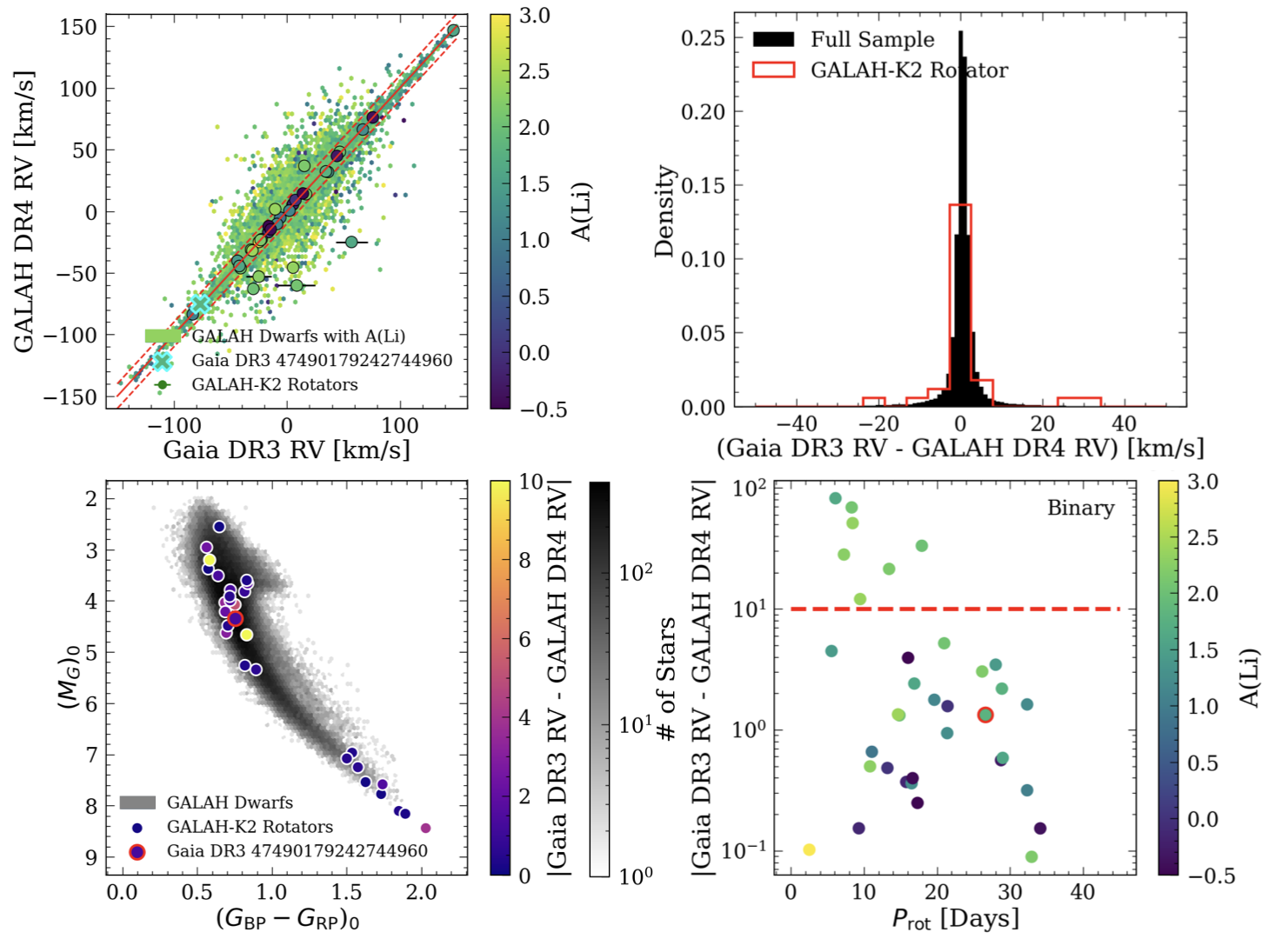}
    \caption{Identifying binaries by combining Gaia DR3 and GALAH DR4 RVs.
    Top left: Gaia RV versus GALAH RV for the full GALAH dwarf sample (background histogram) and the GALAH--K2 rotators (larger points with red outlines), colored by lithium abundances from \cite{Wang2024}. 
    The cross shows the potentially truly young \halpha\ star. 
    The solid red line shows the 1-to-1 line, and the dotted red lines show 10 km/s offset from the 1-to-1 line. 
    We identify a star to be in a binary system if the RV between Gaia and GALAH differs more than 10 km/s.
    Top right: distribution of the RV differences between Gaia and GALAH for the full GALAH dwarf sample (black) and the GALAH--K2 rotators (red).
    19\% of the GALAH--K2 rotators (7 out of 37) are in binary systems, compared to 1.8\% of the full GALAH dwarf sample.
    Bottom left: reddening corrected color-magnitude-diagram (CMD) for the full GALAH dwarf sample (background histogram) and the GALAH--K2 rotators (points, colored by the absolute RV difference between Gaia and GALAH).
    Reddening is corrected using \texttt{dustmap} \citep{Green2018, Green20182}.
    Stars in binary systems appear to be brighter and hotter, as expected.  
    Bottom right: \prot\ versus absolute RV difference between Gaia and GALAH for the GALAH--K2 rotators, colored by lithium abundance measurements. 
    Stars rotating $<$ 10 days are almost all in binary systems.
    }
    \label{fig:fig_galah_gaiarv}
\end{figure*}

Since all GALAH--K2 rotators in our sample only have single RV measurements from GALAH, we compared the RV measurements reported in GALAH DR4 with those reported in Gaia DR3. 
Figure~\ref{fig:fig_galah_gaiarv} left plot shows the comparison between Gaia DR3 and GALAH DR4 RVs colored by lithium abundances.
Compared with the full GALAH dwarf sample (see Figure~\ref{fig:fig_galah_gaiarv} second plot from the left), stars with rotation measurements are more likely to be in binaries (we defined it to be in a binary system when the difference in RV between Gaia and GALAH $>$ 10 km/s; shown as a dotted line in Figure~\ref{fig:fig_galah_gaiarv} right plot).
Quantitatively, 19\% of the GALAH--K2 rotators (7 out of 37) are in binary systems, compared to 1.8\% in the full GALAH dwarf sample, agreeing with studies of young \halpha\ giants and subgiants \citep[e.g.,][]{Jofre2016}.
Faster rotators are also more likely to be in binaries as shown in Figure~\ref{fig:fig_galah_gaiarv} right plot, agreeing with literature that stars rotating $<$ 10 days are most likely in binary systems \citep[e.g.,][]{Simonian2019}. 
Stars that are in binaries appear to be hotter and brighter in the reddening corrected color-magnitude diagram (CMD; see the third plot from the left in Figure~\ref{fig:fig_galah_gaiarv}). 

It is worth noting that the 6 GALAH--K2 binaries identified in Section~\ref{subsec:GALAHAPOGEE} also have RV difference $>$ 10 km/s between Gaia and GALAH. 
After excluding the binaries identified through Gaia, the CMD of the remaining GALAH--K2 rotators now visually matches better with the APOGEE--K2/Kepler rotator sample.
Our final sample contains 41 APOGEE--K2/Kepler stars and 30 GALAH--K2 stars with period measurements, corresponding to period yields of 5\% and 12\% for the APOGEE--K2/Kepler sample and the GALAH--K2 sample, respectively. 
Table~\ref{tab:tab1} shows the catalog description for this final rotating \halpha\ sample.
However, it is worth noting that these yields are not the intrinsic percentage of young \halpha\ stars, since we selected samples that are extremely biased toward single stars for the APOGEE--K2/Kepler sample and stars with lithium measurements for the GALAH--K2 sample. 
\begin{table*}
\centering
\caption{Catalog description of the final sample of 41 APOGEE--K2/Kepler and 30 GALAH--K2 rotators. This table is published in its entirety in a machine-readable format in the online journal.}
\begin{tabular}{crcrcr}
\hline
\hline
Column & Unit & Description \\
\hline
\texttt{sample} &  & Sample name, either ``APOGEE--K2/Kepler'' or ``GALAH--K2''  \\
\texttt{source\_id} &  & Gaia DR3 source ID  \\
\texttt{APOGEE\_id} &  & APOGEE DR17 ID if available \\
\texttt{GALAH\_id} & & GALAH DR3 ID if available \\
\texttt{KIC} &  & KIC ID if available \\
\texttt{EPIC} &  & K2 ID if available \\
\texttt{Prot} & days & rotation period measured \\
\texttt{Age\_STAREVOL} & Gyr & age inferred from STAREVOL \citep[isochrone+gyrochronology;][]{Siess2000, Amard2019} \\
\texttt{e\_Age\_STAREVOL} & Gyr & lower uncertainty on ages inferred from STAREVOL\\
\texttt{E\_Age\_STAREVOL} & Gyr & upper uncertainty on ages inferred from STAREVOL\\
\texttt{Age\_GPgyro} & Gyr & age inferred from GPgyro \citep[gyrochronology;][]{Lu2024} \\
\texttt{e\_Age\_GPgyro} & Gyr & lower uncertainty on ages inferred from GPgyro\\
\texttt{E\_Age\_GPgyro} & Gyr & upper uncertainty on ages inferred from GPgyro\\
\hline
\end{tabular}
\label{tab:tab1}
\end{table*}

In summary, we find 41 APOGEE single stars with rotation periods measured from Kepler and K2 and 37 GALAH stars with period measurements from K2 that also have lithium abundances reported by \cite{Wang2024}.
All these stars have rotation $<$ 40 days, indicating their youth.
We excluded 7 binaries to the best of our ability for the GALAH sample using Gaia and GALAH RV measurements, APOGEE multi-epoch RVs, and double-lined spectroscopy detection (see Section~\ref{subsec:galahgaia}).
We also excluded one pre-main-sequence star with incorrect abundance measurements from the GALAH sample (see Section~\ref{subsec:twostars}). 
This left us with 29 GALAH rotators. 
A discussion on the treatment of binaries can be found in Section~\ref{sec:binaires}.

\section{Results} \label{sec:result}
\subsection{Period Distribution \& 10 Truly Young Candidates}\label{subsec:period}
\begin{figure*}
    \centering
    \includegraphics[width=\linewidth]{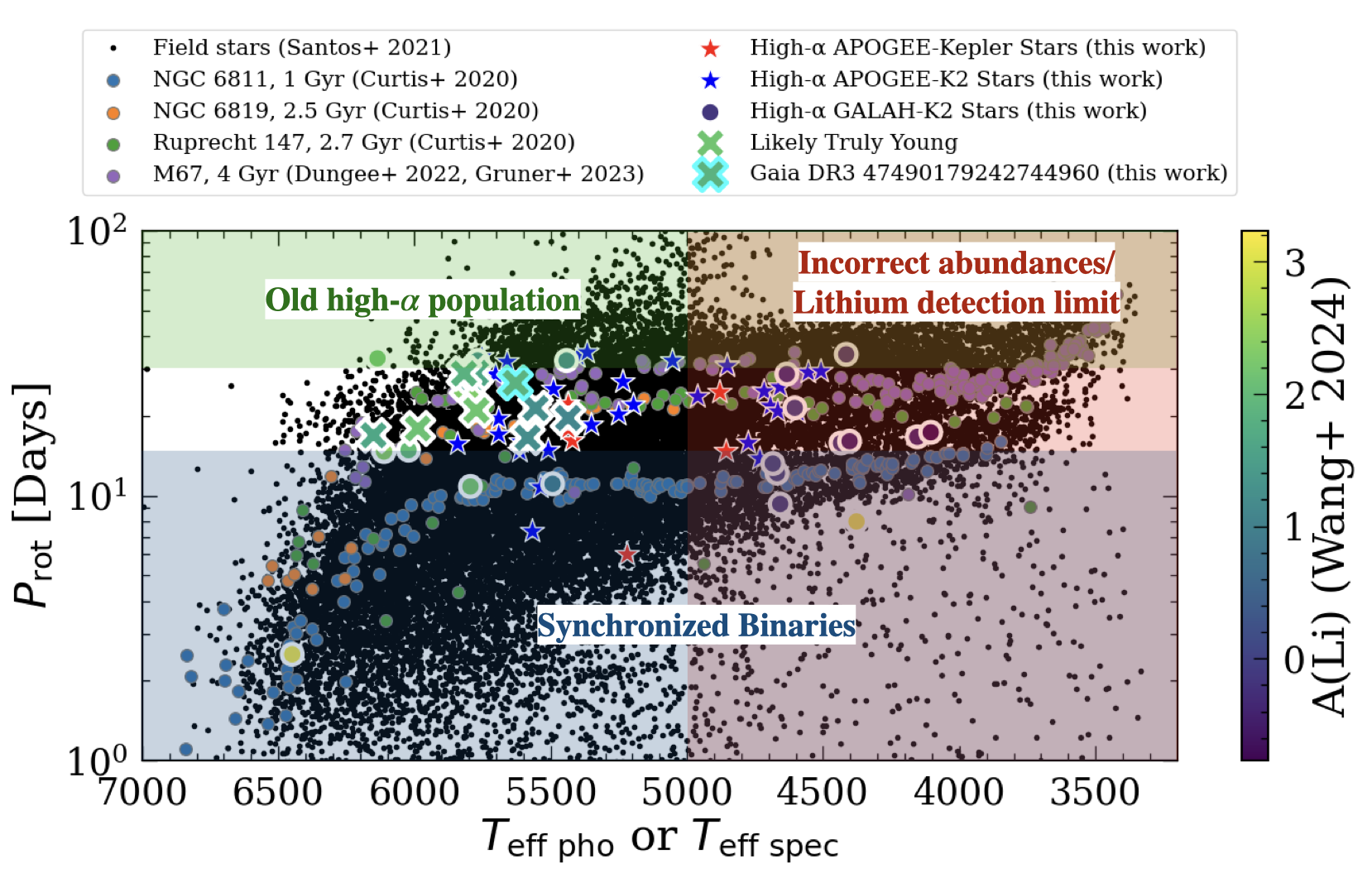}
    \caption{\prot-\teff\ distribution for the 29 GALAH--K2 and the 41 APOGEE--K2/Kepler rotators overplotting on top of the Kepler period detection with RUWE $<$ 1.2 \citep[black points;][]{Santos2021} and a few selected cluster samples between 1-4 Gyr \citep[colored points;][]{Curtis2020, Gruner2023}.
    The GALAH--K2 stars are colored by lithium abundances reported from \cite{Wang2024}.
    The shaded blue region (\prot\ $<$ 15 days) marks the parameter space where binary interaction can likely spin up a main-sequence star, causing an artificially younger prediction for their gyrochronology ages  \citep[e.g.,][]{Meibom2005, Lurie2017, Simonian2019}.
    The shaded red region (\teff\ $<$ 5,000 K) shows the parameter space where lithium measurement is not reliable (same as Figure~\ref{fig:fig1}).
    Lastly, the shaded green region (\prot\ $>$ 30 days) marks the approximate parameter space where we expect to find the old \halpha\ population ($>\sim$ 7 Gyr) according to \texttt{GPgyro}.
    Excluding these regions, 10 stars with lithium measurements from GALAH marked with crosses are prime candidates for truly young \halpha\ stars.
    The star marked with a cross with the cyan outline is the GALAH--K2 single star that is likely truly young (see Section~\ref{subsec:trulyyoung}). 
    }
    \label{fig:fig3}
\end{figure*}

Figure~\ref{fig:fig3} shows the \prot-\teff\ distribution for the 30 GALAH--K2 (excluding 3 binaries identified with APOGEE, 3 that are double-lined spectroscopic binaries, and 1 with RV disagreements $>$ 10 km/s between Gaia DR3 and GALAH) and the 41 APOGEE--K2/Kepler rotators overplotting on the Kepler period detection with Gaia DR3 RUWE $<$ 1.2 \citep[black points;][]{Santos2021} and a few selected cluster samples between 1-4 Gyr \citep[colored points;][]{Curtis2020, Gruner2023}.
The star marked with a cross is the GALAH--K2 single star that could be truly young. 
The spectroscopic temperature is plotted for the GALAH--K2, and APOGE-K2/Kepler stars, and the photometric temperature for the cluster stars is converted from de-reddened $G_{\rm BP}-G_{\rm RP}$ using a polynomial fit taken from \cite{Curtis2020}.
We de-reddened the color using {\tt dustmap} \citep{Green2018, Green20182}.
For the rest of the Kepler stars, we plotted the photometric temperature converted using the same relation from the raw Gaia DR3 $G_{\rm BP}-G_{\rm RP}$.

Looking at this figure, it is clear where the truly young candidates are.
The shaded red region (\teff\ $<$ 5,000 K) shows the parameter space where lithium measurement is not reliable (same as Figure~\ref{fig:fig1}), and thus, cannot provide information on past mass-transfer or merger events.
The shaded green region (\prot\ $>$ 30 days) marks the approximate parameter space where we expect to find the old \halpha\ population ($>\sim$ 6 Gyr to be conservative) according to the gyrochronology model, \texttt{GPgyro}.
The shaded blue region (\prot\ $<$ 15 days) marks the parameter space where most stars are in binaries that are tidally locked, in which their rotation periods and their orbital periods are in sync. 
Stars with \prot\ $<\sim$ 15 days most likely were spun up through binary interactions, causing an artificially younger prediction for their gyrochronology ages  \citep[e.g.,][]{Meibom2005, Lurie2017, Simonian2019}.

Excluding stars in the shaded region, we are left with 10 truly young candidates (including the one mentioned in Section~\ref{subsec:trulyyoung}).
If all of the periods of these stars are spun up via binary interactions, the companions will need to be close in, which would be flagged as having variable RVs with Gaia, which is not the case.
These stars have also not gone through stellar mergers or mass-transfer events as they still retain significant lithium.
As a result, if they are not truly young, an explanation without invoking a past or present stellar binary is needed.

\subsection{K2 Rotators Observed by APOGEE and GALAH}\label{subsec:twostars}
Out of the 12 K2 rotators observed by both APOGEE and GALAH (see Section~\ref{subsec:GALAHAPOGEE}), 4 stars have lithium measurements from GALAH and at least 2 APOGEE visits with their RV measurements agreeing within uncertainty, suggesting they are likely single stars or likely do not have close companions. 
In this section, we will take a closer look at these 4 stars.

\begin{figure}
    \centering
    \includegraphics[width=\columnwidth]{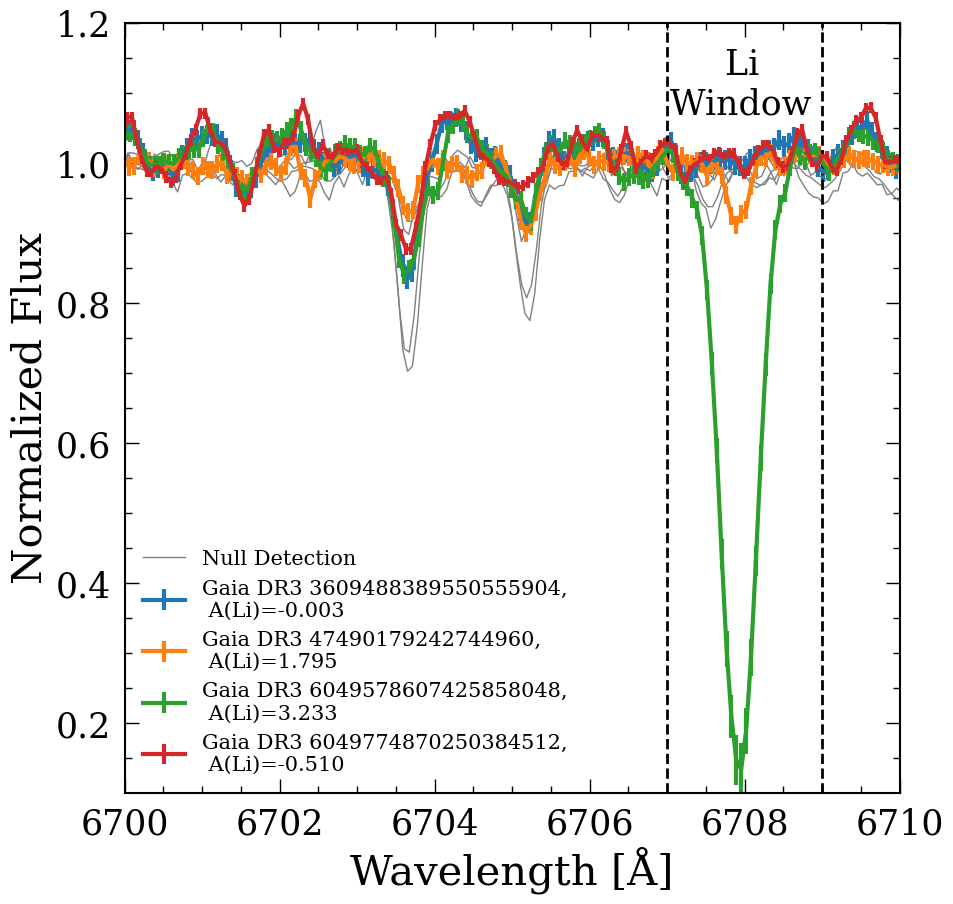}
    \caption{A section of the GALAH red arm spectra covering the lithium window at 6707.814 \AA\ for the 4 likely single GALAH--K2 star with reported lithium from \cite{Wang2024}.
    The spectra for the 4 stars are plotted in colored lines with error bars, and the background grey lines show 4 stars randomly selected that have no lithium detection.
    It is clear that Gaia DR3 47490179242744960 (orange) and Gaia DR3 6049578607425858048 (green) have visible lithium lines, confirming their lithium detection.}
    \label{fig:figA1}
\end{figure}

First, we check the lithium window at 6707.814 \AA\ to ensure that there are indeed reliable lithium detections.
A section of the GALAH red arm spectra covering this line is shown in Figure~\ref{fig:figA1}. 
We overplotted the normalized flux of the 4 GALAH--K2 stars of interest with error bars on top of 4 random stars with no lithium detection according to \cite{Wang2024}. 
It is clear that at least two stars, Gaia DR3 47490179242744960 and Gaia DR3 6049578607425858048 have significant lithium absorption lines.  

The two stars in the $[\mathrm{Fe}/\mathrm{H}]$--\alphafe\ and the \logg-\teff\ parameter space are shown as red outlined points in Figure~\ref{fig:fig1}.
From the Kiel diagram, the star with an extremely high lithium abundance, A(Li) = 3.23, appears to be a pre-main-sequence (PMS) star, agreeing with its fast 8-day rotation period. 
After further checking, it is also part of an 8 Myr young association, the Scorpius–Centaurus association \citep{Luhman2020}.
However, the $[\mathrm{Mg}/\mathrm{Fe}]$ abundance from GALAH DR4 suggests this star is a low-$\alpha$ star that got mistakenly classified as a \halpha\ star due to incorrect abundance measurements, and thus, we exclude this star from our analysis in the rest of this paper and we are left with 29 GALAH--K2 rotators. 
However, for Gaia DR3 47490179242744960, Figure~\ref{fig:fig_galah_apogee} shows the stellar parameters agree within uncertainty between APOGEE and GALAH (green point).  
Further checking abundances and model fit from GALAH DR4 confirms that this is a \halpha\ star.

\subsection{Age Distribution}\label{subsec:agedist}
We used the grid of stellar evolution model by \citet{Amard2019} to estimate the age of each star. This grid is particularly interesting as it provides a rotation period consistently computed along the evolution, with internal rotational mixing, and following the surface magnetized torque by \cite{Matt2015}. We refer to \cite{Amard2019} for a detailed description of the physics of the models.
The grid covers a range of metallicity from $[\mathrm{Fe}/\mathrm{H}]$=-1 to $[\mathrm{Fe}/\mathrm{H}]$=+0.3, and thus encompasses most of the observations. However, the \alphafe\ content is fixed for each metallicity. The values were selected to roughly reproduce the thin disk observations, and therefore do not systematically match the observed value. The grid spreads from 0.2 to 1.5 M$_\odot$, from the pre-main-sequence to the turn-off. To estimate the ages and their uncertainties, we used a maximum likelihood interpolation tool adapted from \citet{Valle2014}, as in \cite{Amard2020}. The interpolator uses the effective temperature, the surface gravity, the metallicity, and the rotation period, and their respective uncertainties, to compare to the theoretical tracks and to provide an age for each star of the sample.

\begin{figure*}
    \centering
    \includegraphics[width=0.32\textwidth]{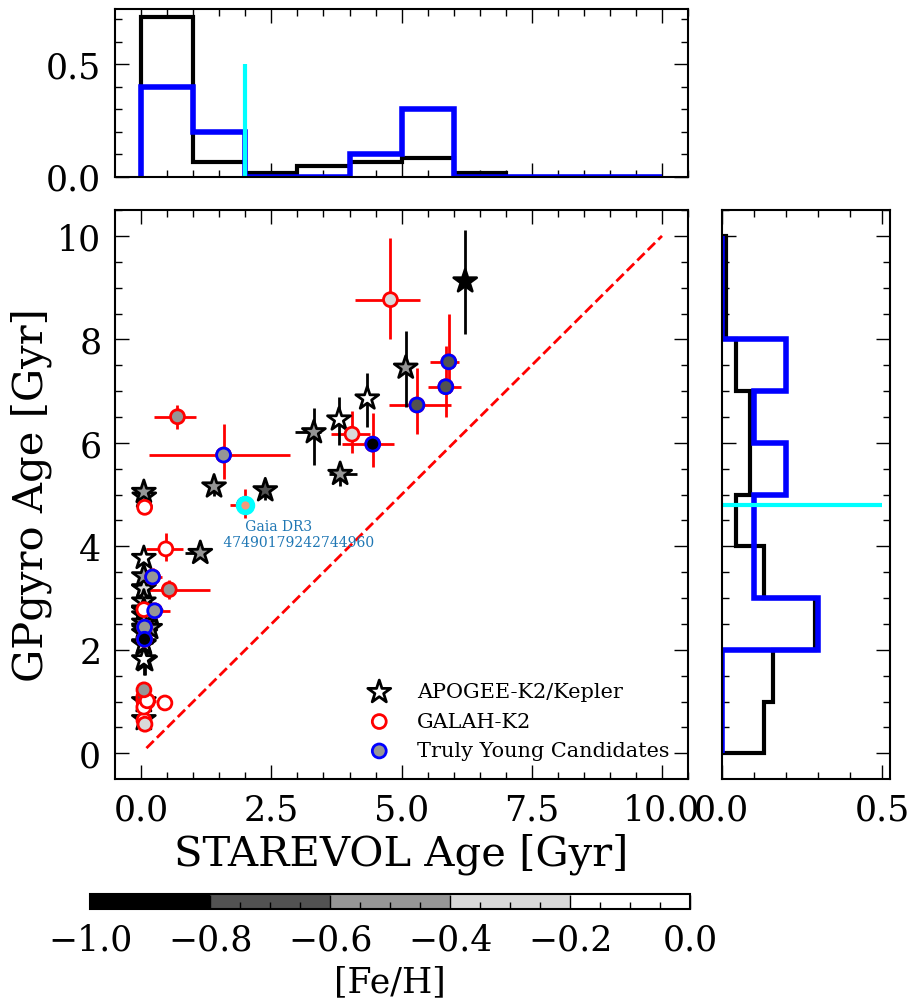}
    \includegraphics[width=0.32\textwidth]{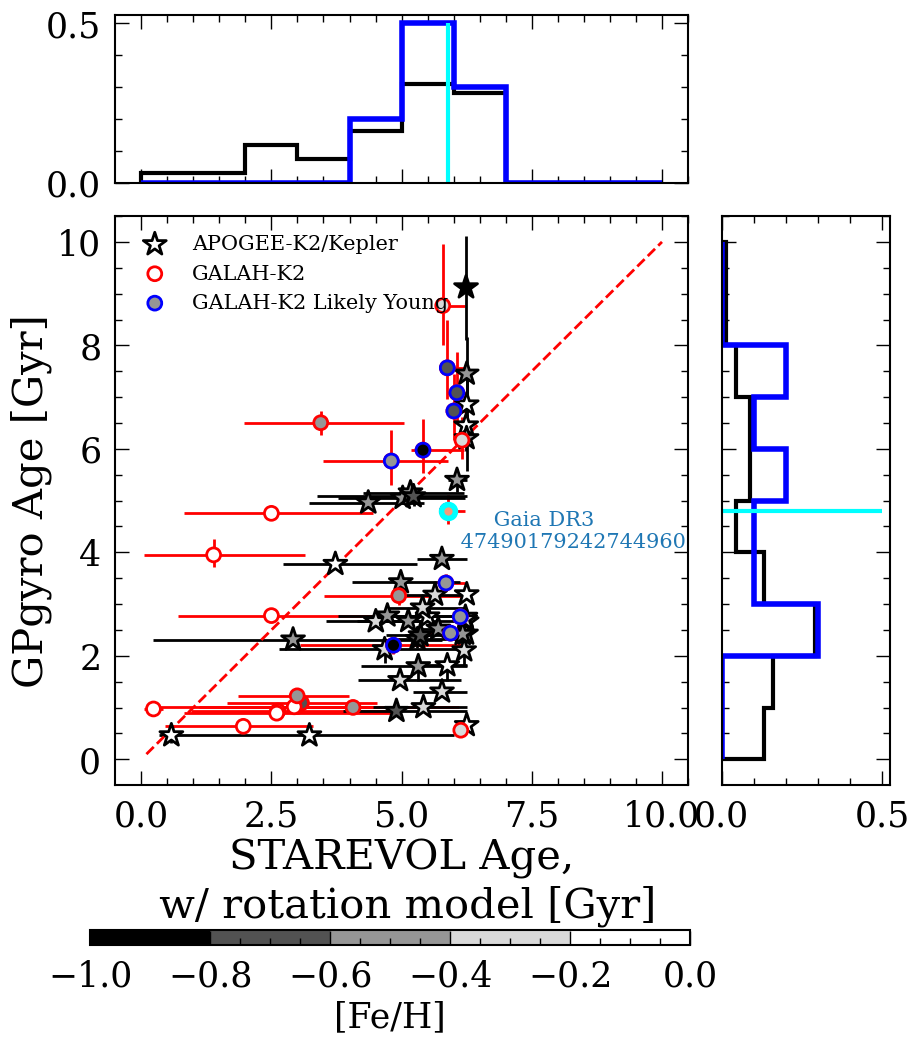}
    \includegraphics[width=0.32\textwidth]{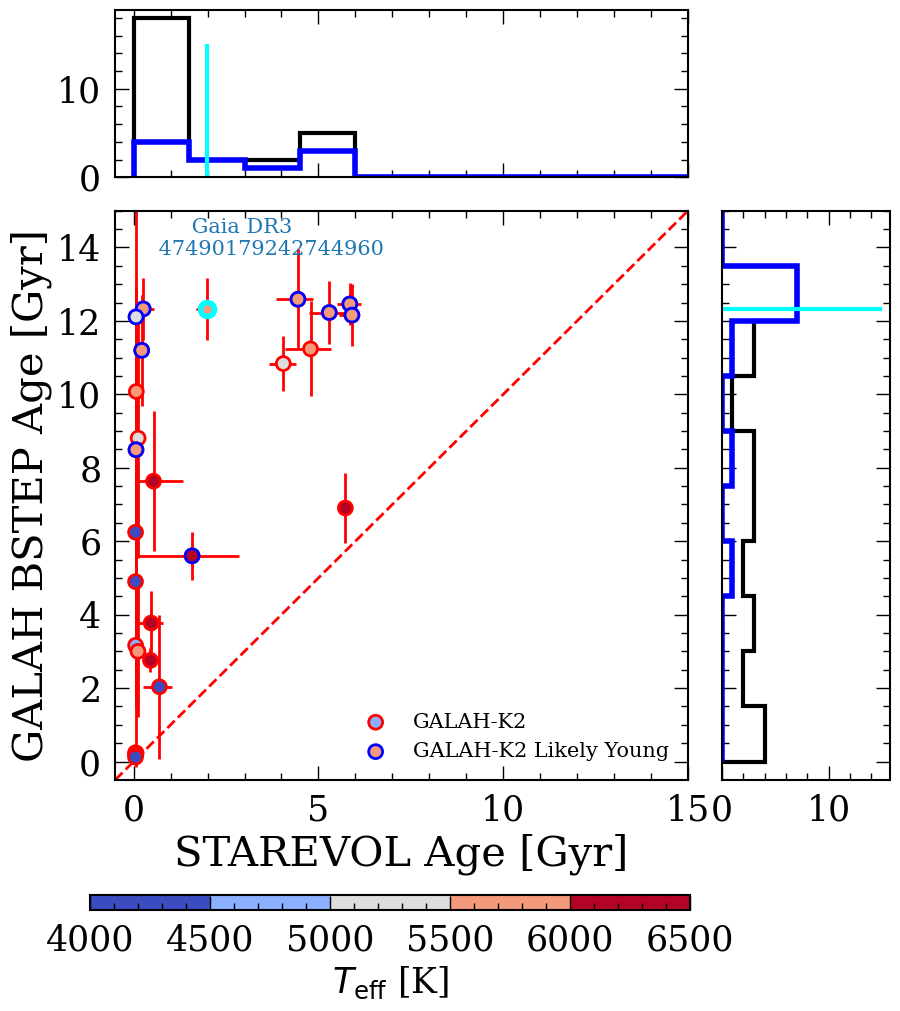}
    
    \caption{Left: Comparison between age inferred from \texttt{STAREVOL} \citep{Siess2000,Amard2019} and \texttt{GPgyro} \citep{Lu2024} for the fast-rotating \halpha\ stars.
    The age distribution for ages inferred from each method is shown on the top and right histograms.
    The blue points and histograms show the ages of the 10 truly young candidate stars (See section~\ref{subsec:period}).
    The cyan point and lines show the age estimates for the possibly truly young \halpha\ star (See section~\ref{subsec:trulyyoung}).
    The bias and variance between the two methods are 2.63 Gyr and 2.00 Gyr, respectively. 
    The pile-up for the \texttt{STAREVOL} ages at $<$ 100 Myr is caused by the limitation of the evaluation grid in that the grid is sparse for stars occupying those specific parameter spaces.
    Middle: Same as left but with the rotation model in \texttt{STAREVOL} turned off.
    Right: Same as left but the comparison between \texttt{STAREVOL} age and \texttt{BSTEP} age provided by GALAH DR3 value added age catalog \citep{Marigo2017, Sharma2018}. }
    \label{fig:age}
\end{figure*}

Empirical gyrochronology ages are calculated using \texttt{GPgyro}\footnote{https://github.com/lyx12311/GPgyro.}\citep{Lu2024}, which is a fully empirical gyrochronology relation calibrated using kinematic ages \citep{Lu2021}, without taking into account metallicity effects.
Figure~\ref{fig:age} left plot shows the comparison between ages inferred from \texttt{STAREVOL} and \texttt{GPgyro}, colored by metallicity derived from APOGEE or GALAH excluding the 9 stars without age estimations from \texttt{STAREVOL}.

In general, there exists a bias of 2.63 Gyr and a variance of 2.00 Gyr between the two methods. 
This can be expected as different stellar evolution models can produce bias on the order of 1 Gyr \citep{SilvaAguirre2017}, and neglecting the effect of metallicity and $\alpha$ can also introduce further bias \citep{Claytor2020, Lu2024}.
The pile-up for the \texttt{STAREVOL} ages at $<$ 100 Myr is caused by the limitation of the evaluation grid in that the grid is sparse for stars occupying those specific parameter spaces.
Nonetheless, most of the stars in our sample have inferred age $<$ 6 Gyr from both \texttt{STAREVOL} and \texttt{GPgyro}.
This means these \halpha\ stars are indeed rotating fast enough to be considered young.
Interestingly, both age-dating methods show a bi-modality in the age distribution, this could indicate two separate channels exist for the formation of the young \halpha\ stars. 
However, it is worth noting that the \texttt{STAREVOL} model also takes into account the rotation periods of stars, and as a result, the two methods are not independent. 
Since most of the stars in our sample have metallicity $<$ 0, we chose to adapt the ages from the \texttt{STAREVOL} model as the \texttt{GPgyro} method is most suitable for stars of solar metallicity. 

Figure~\ref{fig:age} middle plot shows the same as the left plot but with the rotation model turned off.
The pile-up for \texttt{STAREVOL} ages without the rotation model at $\sim$ 6 Gyr reflects the fact that pure isochrone ages have large uncertainty in inferring ages for main-sequence stars, and as a result, the model evaluates the ages to be the mean of the full evaluation time ($\sim$ 13 Gyr).
The right plot shows the same as the left plot but comparing ages for the GALAH--K2 sample inferred from \texttt{STAREVOL} and the Bayesian Stellar Parameter Estimation code \citep[\texttt{BSTEP}, a Bayesian estimate of intrinsic stellar parameters from observed parameters using stellar isochrones, provided by the GALAH DR3 value-added catalog;][]{Marigo2017, Sharma2018}. 

The blue points and histograms show the ages of the 10 truly young candidate stars.
These stars have an average age of 4.8 Gyr according to \texttt{GPgyro} (gyrochronology), 11.1 Gyr according to \texttt{BSTEP} or 5.7 Gyr according to \texttt{STAREVOL} without the built-in spin-down model (isochrone), and 2.6 Gyr according to \texttt{STAREVOL} (isochrone + gyrochronology).
On average, the age estimates except \texttt{BSTEP} for these stars agree with their truly young nature. 

The cyan point and lines in both plots show the age estimate for the one truly young \halpha\ candidate star that we did a deeper analysis of (see Section~\ref{subsec:trulyyoung}).
This star is estimated to be 4.8$^{+0.31}_{-0.25}$ Gyr based on \texttt{GPgyro}, 12.32 $\pm$  0.85 Gyr based on \texttt{BSTEP} or 5.88 $^{+0.08}_{-0.34}$ based on \texttt{STAREVOL} without the built-in spin-down model (isochrone), and 1.98$^{+0.12}_{-0.28}$ Gyr based on \texttt{STAREVOL} (isochrone + gyrochronology). 
Since these stars are main-sequence stars, pure isochrone models most likely will over-predict ages for these stars \citep[e.g.,][]{Byrom2024}.
However, it is worth pointing out that this star has a high extinction value (E(B-V)=0.67 mag), which is not accounted for fully by the \texttt{BSTEP} model, which reports an E(B-V) value of 0.29 mag.
Since this is a main-sequence G dwarf, age inferred from gyrochronology is likely to be more accurate, but we do recognize that it is unclear what causes the age discrepancy. 

\subsection{Kinematic Distribution}\label{subsec:kinematic}
\begin{figure*}
    \centering
    \includegraphics[width=\linewidth]{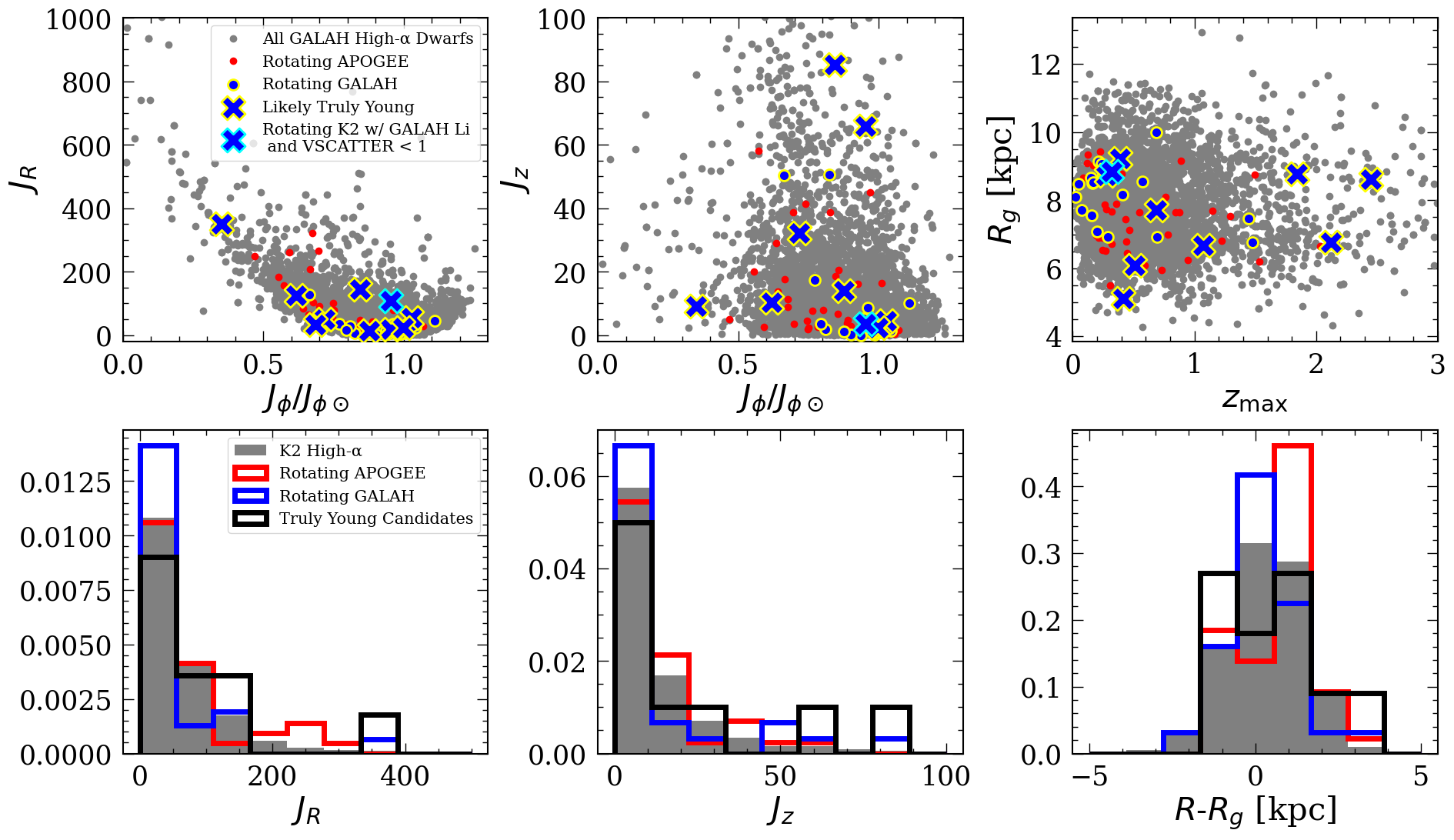}
    \caption{Top row: the angular momentum, $J_\phi$, versus the radial action, $J_R$ (left); $J_\phi$, versus the vertical action, $J_z$ (middle); and the maximum vertical extent, $z_{\rm max}$, versus the guiding radius, $R_g$ (right) of all the \halpha\ dwarf stars identified by GALAH (grey) and the rotating APOGEE--K2/Kepler and GALAH--K2 \halpha\ star sample (blue and red as indicated in the legends).
    The blue crosses show the 10 truly young \halpha\ candidates and the cross with the cyan outline shows Gaia DR3 47490179242744960, the one potential genuinely young \halpha\ star identified in Section~\ref{subsec:trulyyoung}
    Bottom row: normalized histograms $J_R$ (left), $J_z$ (middle), and the difference between the current radius and guiding radius ($R-R_g$; right) for the \halpha\ dwarf stars identified by GALAH (grey), the APOGEE--K2/Kepler rotating \halpha\ stars (red), the GALAH--K2 rotating \halpha\ stars with lithium measurements (blue), and the 10 truly young candidates (black).}
    \label{fig:fig2}
\end{figure*}

Figure~\ref{fig:fig2} top plots show the kinematic distribution of all the \halpha\ dwarf stars identified by GALAH (grey) and the ones with period detected (blue and red as indicated in the legends).
The blue crosses show the 10 truly young \halpha\ candidates, and the blue cross with the cyan outline shows Gaia DR3 47490179242744960, the one potential genuinely young \halpha\ star identified in Section~\ref{subsec:trulyyoung}.
The bottom plots show the normalized histogram of the \halpha\ GALAH dwarf stars (grey), the rotating APOGEE--K2/Kepler \halpha\ stars (red), the rotating GALAH--K2 \halpha\ stars (blue), and the 10 truly young \halpha\ candidates (blue).
All kinematic properties including Galactocentric radius ($R$) are calculated from Gaia DR3 measurements (RA, Dec., parallax, proper motions, and RV) \citep{Gaiadr3} by transforming from the solar system barycentric ICRS reference frame to the Galactocentric Cartesian and cylindrical coordinates using \texttt{astropy} \citep{astropy:2013, astropy:2018, astropy2022} using updated solar motion parameters from \citet{Hunt:2022}.
The actions ${\bf J} = (J_R, J_\phi, J_z)$, guiding radius ($R_g$), and maximum vertical extension ($z_{\rm max}$) are calculated using the \texttt{MilkyWayPotential2022} in \texttt{gala} \citep{gala2017}, with the additional constraint of having a circular velocity at the solar position to be 229 km/s \citep{Eilers2019}. 
We then compute actions using the `St\"{a}ckel Fudge' \citep{Binney2012, Sanders2012} as implemented in \texttt{galpy} \citep{galpy}.

The kinematic distribution of the APOGEE--K2/Kepler \halpha\ stars and the 10 truly young candidate stars agree with the full underlying sample, which is a mix of thin and thick-disk stars.
What is interesting is that the full sample of rotating GALAH--K2 sample with lithium measurements mostly exhibits thin-disk kinematics --- most are on highly circular orbits in the solar neighborhood, including Gaia DR3 47490179242744960, which is on a thin-disk orbit ($z_{\rm max}< 0.5$ kpc) that is slightly eccentric.
However, stars on thin-disk orbits do not necessarily mean they are young and it is possible that some low-$\alpha$ stars can be misclassified as \halpha\ due to incorrect abundance measurements.
\halpha\ stars on thin-disk orbits do exist in both simulations and observations, this is especially common for stars with $[\mathrm{Fe}/\mathrm{H}] > -1$, which is when the rotationally supported stellar disk has started to form \citep{Belokurov2022}.

\subsection{Abundance Distribution}\label{subsec:abundance}
\begin{figure*}
    \centering
    \includegraphics[width=\linewidth]{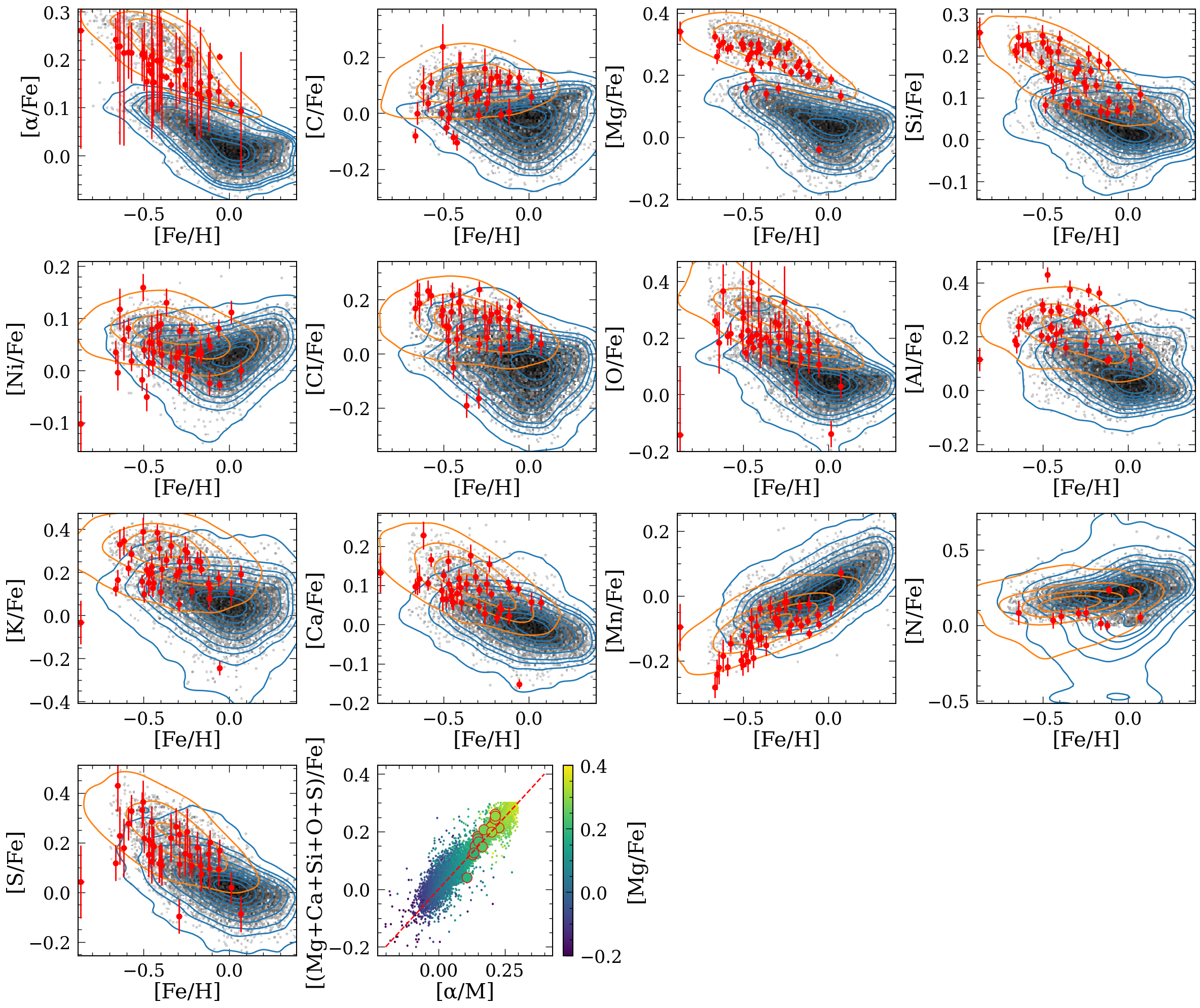}
    \caption{abundance distribution for the 41 APOGEE--K2/Kepler \halpha\ rotators that are potentially truly young (red points; selection see Figure~\ref{fig:fig3}]) for elements that are identified as most reliable (C, Mg, Si, Fe, Ni), reliable (C I, O, Al, K, Ca, Mn), and the less reliable (N, S) in APOGEE dwarfs. 
    The background grey histograms show the full APOGEE--K2/Kepler abundance distributions, and the orange and blue contours show the kernel density estimate (KDE) for the high- (orange) and low-$\alpha$ (blue) stars, identified with the red dashed line in Figure~\ref{fig:fig1}.
    For each element, the element flags reported by ASPCAP are set to 0 to ensure the measurements are reliable. 
    The last plot shows the averaged $\alpha$ abundance calculated from the individual $\alpha$ elements compared with the reported $[\mathrm{\alpha}/\mathrm{M}]$ values from ASPCAP colored by the reported $[\mathrm{Mg}/\mathrm{Fe}]$ values for the full sample (background histogram)  and the APOGEE--K2/Kepler rotators (points).}
    \label{fig:fig4}
\end{figure*}

Figure~\ref{fig:fig4} shows the abundance distribution for the 41 APOGEE--K2/Kepler \halpha\ rotators (red points) for elements that are identified as most reliable (C, Mg, Si, Fe, Ni), reliable (C I, O, Al, K, Ca, Mn), and less reliable (N, S) in APOGEE dwarfs\footnote{see \url{https://www.sdss4.org/dr17/irspec/abundances/} for more detail of reliable abundances in APOGEE.}. 
The background grey histograms show the full APOGEE--K2/Kepler abundance distributions, and the orange and blue contours show the kernel density estimate (KDE) for the high- (orange) and low-$\alpha$ (blue) stars, identified with the red dashed line in Figure~\ref{fig:fig1}.
The KDEs are estimated using the \texttt{seaborn.kdeplot} function \citep{seaborn}.
The last plot shows the averaged $\alpha$ abundance calculated from the individual $\alpha$ elements compared with the reported $[\mathrm{\alpha}/\mathrm{M}]$ values from ASPCAP.
In general, the averaged $\alpha$ abundances agree well with the reported $[\mathrm{\alpha}/\mathrm{M}]$ values.
It is worth noting that one star has a high $\alpha$ abundance measurement but a low $[\mathrm{Mg}/\mathrm{Fe}]$ measurement.
The best-fit model for this star agrees well with the actual spectra but it has a 5-day rotation period, with a $v$sin$i$ value of 7.5 km/s.
As a result, this star could be one that is misidentified as \halpha\ stars due to an incorrect global $[\mathrm{\alpha}/\mathrm{M}]$ fit. 

Comparing the APOGEE--K2/Kepler \halpha\ rotators with the full sample, their individual abundances agree well with those of \halpha\ stars without rotation detections, as also seen by other studies of the young \halpha\ stars \citep[e.g.,][]{Martig2015}. 

\begin{figure*}
    \centering
    \includegraphics[width=\linewidth]{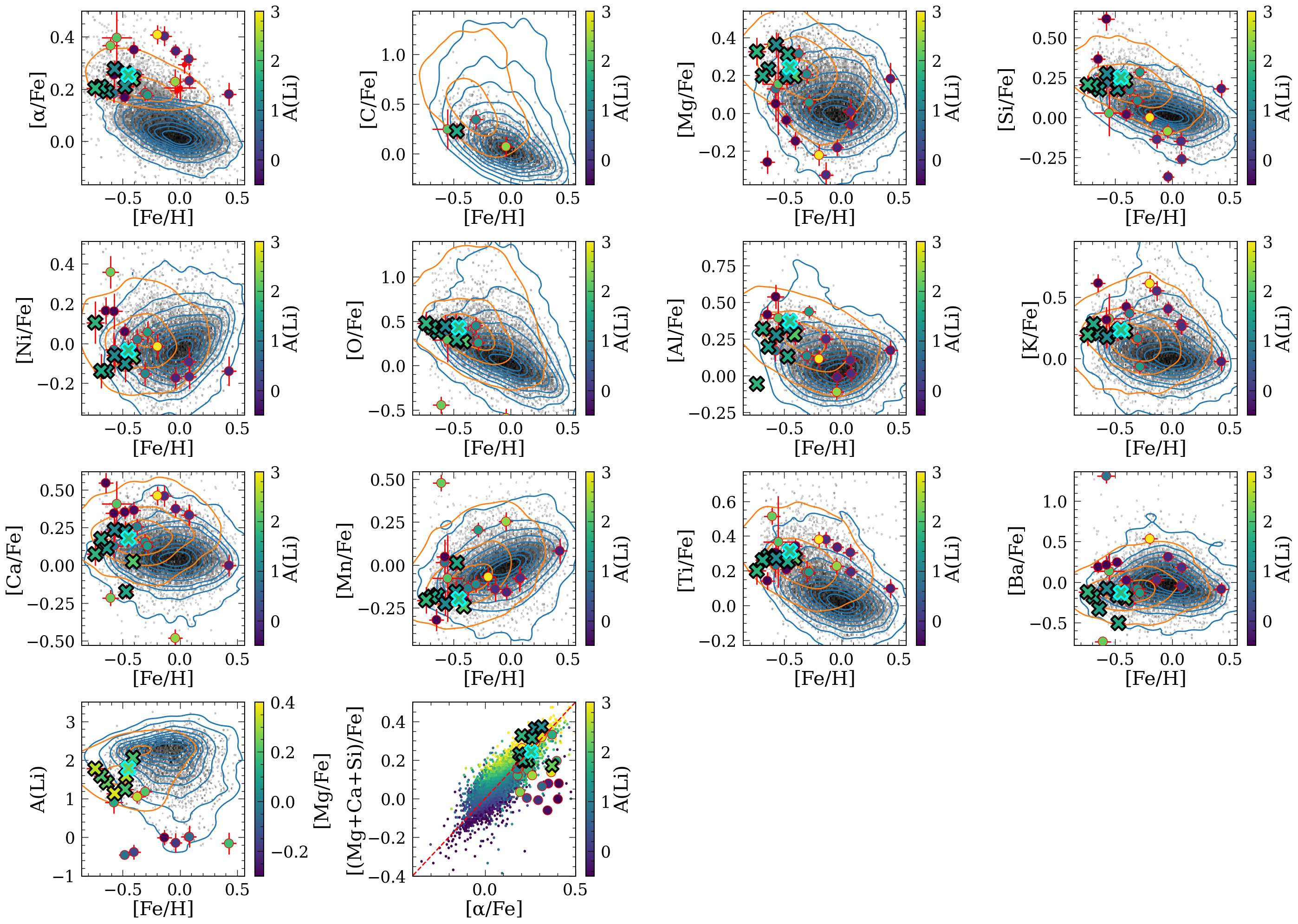}
    \caption{Same as Figure~\ref{fig:fig4} but with the 10 GALAH--K2 potentially young candidate shown in crosses. The cross with a cyan outline represents the likely young \halpha\ rotator mentioned in Section~\ref{subsec:trulyyoung}.
    The 10 candidates visually clump in chemical space, and this could be due to our narrow selection criteria or suggesting their common origin.}
    \label{fig:fig5}
\end{figure*}

Figure~\ref{fig:fig5} shows the same as Figure~\ref{fig:fig4} but for the 29 GALAH--K2 stars. 
The points are colored by lithium abundance measurements, and the crosses are the 10 truly young candidates, and the cross with a cyan outline identifies the candidate that we performed a deeper analysis of (see Section~\ref{subsec:twostars}).
Interestingly, The 10 prime candidates visually clump in chemical space, and this could be due to our narrow selection criteria or suggesting their common origin.
The \alphafe\ reported from GALAH DR3 does not match well with the averaged individual $\alpha$ element abundances (Mg, Ca, Si)\footnote{We did not include oxygen abundance in the calculation as not many stars have reliable oxygen abundance reported.}, especially for the stars that are $<$ 5,000 K and have low lithium abundance measurements.
It is worth pointing out that these stars are split into two groups in both \teff-\logg (see Figure~\ref{fig:fig1}) and abundance space --- the first group with lithium abundance and \teff $>$ 5000 K, and the other group with minimal or no significant lithium detection and \teff $<$ 5000 K. 
This could indicate two formation channels for the young \halpha\ stars. 
However, stars with temperature measurements $<$ 5000 K also likely have inaccurate inferred abundances.
Since the $\alpha$ abundances do not agree between the global fit and the averaged individual $\alpha$ elements, the grouping in abundance space is likely caused by systematic.
However, improving individual abundance measurements for low-mass stars is needed to further test this hypothesis. 

\subsection{A truly young \halpha\ star}\label{subsec:trulyyoung}
For Gaia DR3 47490179242744960, with a lithium measurement A(Li) = 1.79, an estimation of age of 1.98$^{+0.12}_{-0.28}$ Gyr (see Section~\ref{subsec:agedist}), the stellar parameters agree well between APOGEE and GALAH, both surveys confirm that it is a \halpha\ star and its reported \alphafe\ value agrees with the average of all individual $\alpha$ abundances (Mg, Ca, Si, O) reported in GALAH (see Figure~\ref{fig:fig5} last subplot).
In this section, we will look closer at this star and make a case that it is truly young.

\subsubsection{Independent detailed abundance validation}
To further validate the stellar parameters and high $\alpha$ abundances, we analyze the GALAH DR3 spectrum using Brussels Automatic Code for Characterizing High accUracy Spectra \citep[BACCHUS,][]{BACCHUS}.  BACCHUS is a spectral synthesis and fitting tool designed for high-resolution data. 
It generates synthetic spectra using the radiative transfer code TURBOSPECTRUM \citep{Plez2012} adopting the MARCS model atmosphere grid \citep{Gustafsson2008} and assuming one-dimensional local thermodynamic equilibrium (1D LTE).  
We adopt version 5 of the Gaia-ESO linelist for our atomic transition data \citep{Heiter2021} and combine molecular transition data from numerous sources: CH from \citealt{Masseron2014}, C2, CN, OH, and MgH from T. Masseron, private communication, SiH from \citealt{Kurucz1992}, and TiO, FeH, and ZrO from B. Pelz, private communication. We point readers to e.g., \citet{Hawkins2016, Hayes2022}, and the BACCHUS manual for a detailed description of BACCHUS and \citet{Nelson2021} for a complete list of the atomic data used in this analysis.  We generate a model atmosphere adopting the GALAH DR3-reported stellar parameters and confirm that they satisfy Fe ionization-excitation balance, an indicator of suitable $\rm T_{eff}$, log g, and microturbulence.  Fe, Mg, Si, and Ca abundances are then determined using the GALAH DR3-optimized line selection.  We confirm that this star is $\alpha$--rich, with $[\mathrm{Mg}/\mathrm{Fe}]$, $[\mathrm{Ca}/\mathrm{Fe}]$, $[\mathrm{Si}/\mathrm{Fe}]$ of 0.37$\pm$0.08, 0.31$\pm$0.11, and 0.20$\pm$0.11 dex which corresponds to an average \alphafe\ of 0.29$\pm$0.07 dex.

\subsubsection{RV measurements and binary analysis}
\begin{figure}
    \centering
    \includegraphics[width=\columnwidth]{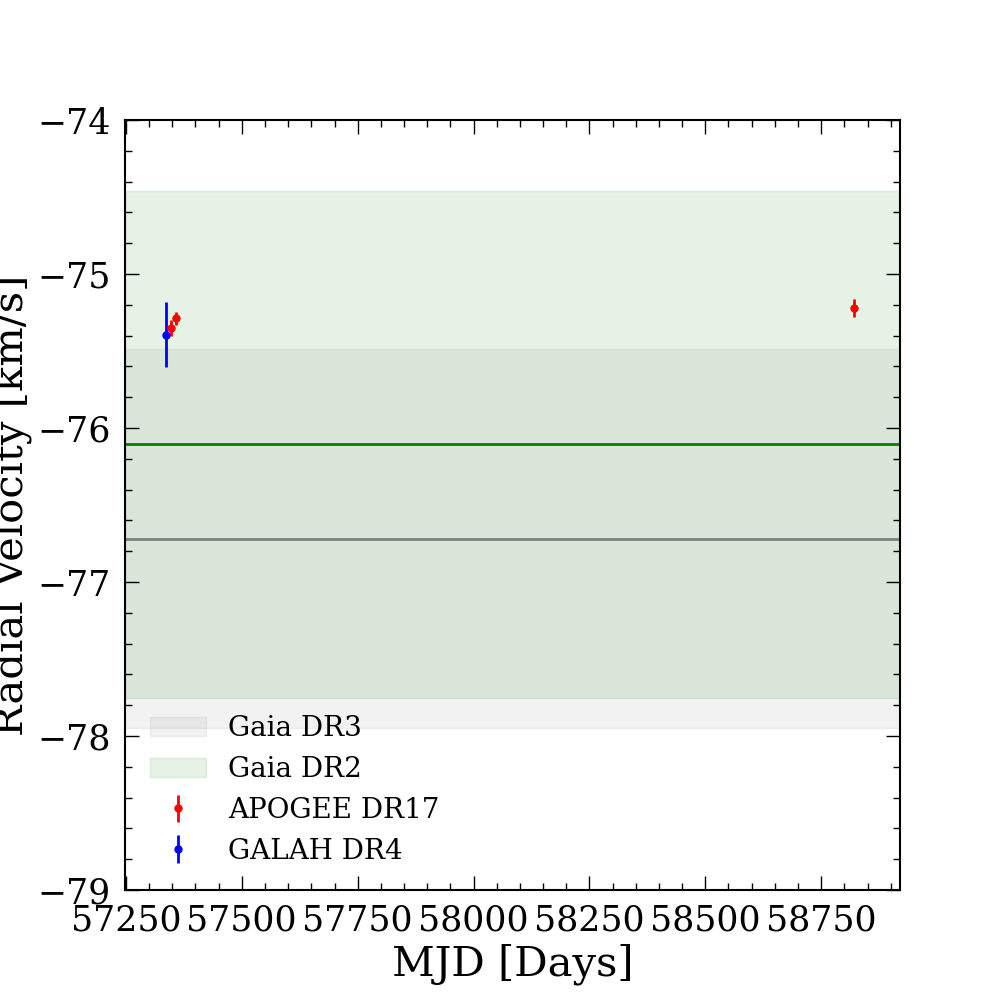}
    \caption{RV measurements from APOGEE (red), GALAH (blue), Gaia DR2 (green shaded area), and Gaia DR3 (grey shaded area).
    The \texttt{The Joker} result shows it is unlikely the star's rotation period is affected by a binary companion.}
    \label{fig:RV_onestar}
\end{figure}

This likely truly young star has three RV measurements from APOGEE, with an average value of -75.3 km/s, an average RV uncertainty of 50 m/s, and an RV scatter between the three measurements of 40 m/s, which agree within uncertainty, suggesting it is likely a single star. 
Figure~\ref{fig:RV_onestar} shows the RV measurements from APOGEE (red), GALAH (blue), Gaia DR2 (green shaded area), and Gaia DR3 (grey shaded area).
Using the \texttt{The Joker} \citep{Joker}, we conclude that if this star is in a binary system, it is unlikely to be a close binary.
That is, using posterior samples over binary orbital parameters generated with The Joker, the posterior probability of this potential system having an RV semi-amplitude $>$1 km/s and an orbital period $<$100 days is only 3\%.
This means it is unlikely the star's rotation period is affected by a binary companion since a companion with an orbital period $>$100 days should not significantly affect the spin-down of the primary star unless the companion is massive.

\subsubsection{Planet engulfment analysis}
\begin{figure*}
    \centering
    \includegraphics[width=\textwidth]{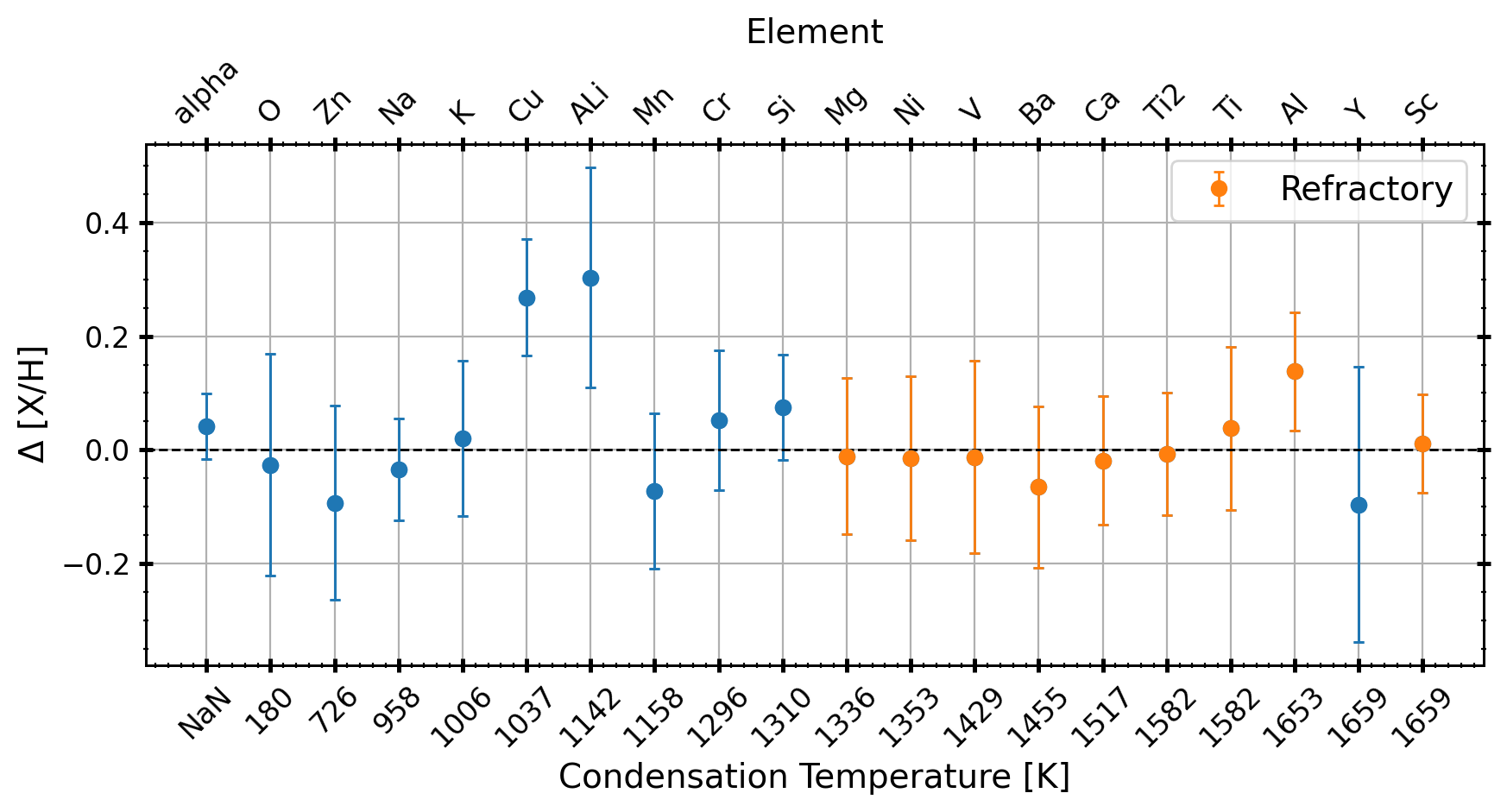}
    \caption{Difference in abundances between Gaia DR3 47490179242744960 and 10 reference stars sorted by condensation temperature. The orange points indicate refractory elements. All abundance measurements were obtained from GALAH DR3, except for lithium for which a 3D-NLTE value from \citet{Wang2024} was used. The lack of significant difference in the refractory elements suggests an absence of planet engulfment.}
    \label{fig:element_diff}
\end{figure*}

The significant lithium line from GALAH, shown in Figure~\ref{fig:figA1} suggests this likely truly young star has not gone through a stellar mass-transfer or merger event, since lithium only exists in the surface layer of main-sequence stars and can be easily destroyed from a stellar merger/mass-transfer event \citep[e.g.,][]{Ryan2001, Pinsonneault2002, Ryan2002}.
However, unlike stellar mergers, a planet engulfment event can spin up the star without destroying its lithium. 
As a result, to suggest the star is likely a truly young \halpha\ star, we also need to exclude the possibility of past planet engulfment events. 

To investigate evidence of planet engulfment in Gaia DR3 47490179242744960, we analyzed chemical abundances, comparing both volatile and refractory elements as a function of condensation temperature, $T_C$. Condensation temperature is the temperature at which an element condenses from gaseous to solid phase, and is often used to study dust and rock formation. Rocky planets would contain an excess of higher condensation temperature elements; therefore, the presence of these elements in the stellar atmosphere at late stages would suggest an engulfment event \citep{Ramirez2009, Teske2016, Melendez2017, Spina2021}.

Figure~\ref{fig:element_diff} compares 20 elemental abundances for Gaia DR3 47490179242744960 with averaged abundances from doppelg{\"a}ngers stars in GALAH. Reference stars were selected based on similar \teff, \logg, $[\mathrm{Fe}/\mathrm{H}]$, and $[\mathrm{Mg}/\mathrm{Fe}]$, while also accounting for each parameter's error. GALAH flags \texttt{flag\_sp} and \texttt{flag\_fe\_h} were enforced to ensure high quality of abundance measurements. Of the total 588,571 stars, 407,737 satisfied the quality cuts, of which 407,519 had a measurement available for \teff, \logg, $[\mathrm{Fe}/\mathrm{H}]$, and $[\mathrm{Mg}/\mathrm{Fe}]$ and its associated error, of which 171,711 had a measurement available for A(Li) as well. The distance in four-dimensional space was minimized to select 10 reference stars for comparison. The mean difference in \teff, \logg, $[\mathrm{Fe}/\mathrm{H}]$, and $[\mathrm{Mg}/\mathrm{Fe}]$ between Gaia DR3 47490179242744960 with its 10 reference stars is 1.6 K, 0.01 dex, 0.002 dex, and 0.01 dex, respectively.

All abundances were measured in GALAH DR3 \citep{Galahdr3}, except for lithium ($A\textrm{(Li)}$) for which a 3D non-local thermodynamic equilibrium (NLTE) value was obtained from \cite{Wang2024} for higher precision. Figure~\ref{fig:element_diff} shows the differences between Gaia DR3 47490179242744960 and its reference stars as a function of elements, sorted by condensation temperature. The orange points indicate refractory elements, which have higher condensation temperatures. 
Error bars were calculated as follows --- the uncertainty in the difference represents the two errors added in quadrature, where the error for the reference star is the average error for all 10 twin stars. 
Differential abundance analysis shows no indication of planet engulfment based on the distribution of refractory elements, and the excess of lithium in this star compared to its reference stars could also indicate that this star is younger than stars of similar $[\mathrm{Fe}/\mathrm{H}]$ and $[\mathrm{Mg}/\mathrm{Fe}]$ as lithium depletes with age on the main-sequence \citep[e.g,][]{Skumanich1972}. 
We also investigated the over-abundance of Cu with the GALAH best-fit spectra and found that it is over-predicted as the Cu absorption lines overlap with the diffuse interstellar bands and interstellar potassium.

\section{Treatment of Binaries}\label{sec:binaires}
For this study, we excluded all binaries to the best of our ability.
However, this is a conservative approach as only close-in binaries should spin up the surface of a star to less than 30 days. 
Main-sequence stars with period measurements $>$ 15 days are likely not in close-in or synchronized binaries as we do not see such a population in open clusters or field star studies.
As a result, the 10 GALAH-K2 G dwarfs with lithium measurements that are rotating between 15-30 days are still most likely genuinely young \halpha\ dwarf stars even with a lack of binary information.
Future studies can relax the binary selection criteria in search of a larger population of intermediate-age \halpha\ dwarf stars.

However, it is unlikely but still possible that tidal synchronization can extend to longer periods for these \halpha\ dwarf stars.
For example, we find one APOGEE \halpha\ star with a period measurement of 18.73 days reported by \cite{Santos2021} that is in a binary system with a 12.4-day orbital period based on 20 APOGEE RV measurements. 
It is unclear how much the rotation of this star is affected by its binary companion, but this seems to be the only edge case in the 20 \halpha\ APOGEE dwarf stars with RUWE $<$ 1.2 and periods $>$ 15 days reported from \cite{Santos2021}. 

\section{Conclusion \& Future Work}
We investigated the existence of genuinely young \halpha\ dwarf stars using photometry from Kepler and K2, and spectroscopy from GALAH and APOGEE. 

After excluding binaries and incorrect abundance measurements in the best of our ability, we find 41 APOGEE single stars with rotation periods measured from Kepler and K2 and 29 GALAH stars with period measurements from K2 that also have lithium abundances reported by \cite{Wang2024} (see Section~\ref{sec:dataselection}).
All these stars have rotation $<$ 40 days, indicating their youth.
Using \prot-\teff\ diagram, we selected 10 prime candidates for being truly young \halpha\ stars.
We suggested if these stars are old, an explanation without invoking stellar binary interaction is needed (see Section~\ref{subsec:period}).
Stars with lithium measurements have ages measurements $<$ 7.5 Gyr using \texttt{STAREVOL}, which is a stellar evolution model with a built-in gyrochronology relation (see Section~\ref{subsec:agedist}).
These stars exhibit mostly thin-disk kinematic (see Section~\ref{subsec:kinematic}), and chemical abundances (see Section~\ref{subsec:abundance}) similar to the rest of \halpha\ stars. 
The 10 prime candidates visually clump in chemical space, and this could be due to our narrow selection criteria or suggesting their common origin.

Within these candidate stars, Gaia 4749017924274496 shows evidence for being a genuinely young \halpha\ star based on the following criteria:
\begin{itemize}
    \item It has an age of 1.98$^{+0.12}_{-0.28}$ Gyr, determined using \texttt{STAREVOL}, a stellar model with a built-in gyrochronology model.
    \item It exhibits planer, thin-disk kinematic, further suggesting its likely young nature.
    However, it is worth pointing out that stars that are on thin-disk orbits are not all young.
    \item It has a significant lithium measurement, A(Li) = 1.79, slightly higher than stars with agreeing $[\mathrm{Fe}/\mathrm{H}]$, $[\mathrm{Mg}/\mathrm{H}]$, \teff, and \logg, suggesting it has not gone through stellar mass transfer or merger \citep[e.g.,][]{Ryan2001, Pinsonneault2002, Ryan2002}.
    \item It is likely single, determined using \texttt{The Joker} by combining RV measurements from APOGEE, GALAH, and Gaia, suggesting its rotation period has not been affected by interaction with a stellar binary companion.  
    \item It likely has not gone through planet engulfment, suggesting the period measurement is not affected by a previously existing planet. 
\end{itemize}
However, it is worth pointing out that an isochrone model without taking into rotation, \texttt{BSTEP}, suggests this star is $>$ 12 Gyr, disagreeing with both \texttt{STAREVOL} and \texttt{GPgyro}, an empirical gyrochronology model.
However, this is not surprising as isochrone fitting cannot provide useful constraints below the field turnoff. 
This further suggests the importance of using gyrochronology for this study, without it, no young \halpha\ dwarf stars can be discovered or confirmed.

Future studies include optical spectroscopy follow-up on the APOGEE--K2/Kepler sample for lithium measurements and multi-epoch RV follow-up on the 10 GALAH--K2 candidate for excluding close-by binary companions, as well as more overlapping optical spectroscopy data for stars with rotation period measurements to extend this study further down the main-sequence.

\section{Acknowledgments}
Y.L. wants to thank the discussion and input from the Nearby University Group at CCA, as well as those from Cecilia Mateu, James Johnson, Diogo Souto, Katia Cunha, Phillip Cargile, and Sean Matt.

This work has made use of data from the European Space Agency (ESA)
mission Gaia,\footnote{\url{https://www.cosmos.esa.int/gaia}} processed by
the Gaia Data Processing and Analysis Consortium (DPAC).\footnote{\url{https://www.cosmos.esa.int/web/gaia/dpac/consortium}} Funding
for the DPAC has been provided by national institutions, in particular
the institutions participating in the Gaia Multilateral Agreement.
This research also made use of public auxiliary data provided by ESA/Gaia/DPAC/CU5 and prepared by Carine Babusiaux. 
L.A. acknowledges support from the Centre National des Etudes Spatiales (CNES) through a PLATO/AIM grant.
S.B. acknowledges support from the Australian Research Council under grant numbers CE170100013 and DE240100150.

% exoplanet archieve
%This research has made use of the NASA Exoplanet Archive, which is operated by the California Institute of Technology, under contract with the National Aeronautics and Space Administration under the Exoplanet Exploration Program \citep{NEA}.

% OSG
%This research was done using services provided by the OSG Consortium \citep{OSG1,OSG2}, which is supported by the National Science Foundation awards \#2030508 and \#1836650.

% SIMBAD, Vizier, ADS
This research has also made use of NASA's Astrophysics Data System, 
and the VizieR \citep{vizier} and SIMBAD \citep{simbad} databases, 
operated at CDS, Strasbourg, France.


\vspace{5mm}
\facilities{Gaia, Kepler, K2, APOGEE, GALAH}


\software{  \texttt{Astropy} \citep{astropy:2013, astropy:2018, astropy2022},
\texttt{lightkurve} \citep{lightkurve2018},
            \texttt{Matplotlib} \citep{matplotlib}, 
            \texttt{NumPy} \citep{Numpy}, 
            \texttt{Pandas} \citep{pandas}, 
            \texttt{gala} \citep{gala2017},
            \texttt{galpy} \citep{galpy},
            \texttt{seaborn} \citep{seaborn},
            \texttt{The Joker} \citep{Joker}}



%\appendix

%\section{Appendix information}

\bibliography{sample631}{}
\bibliographystyle{aasjournal}



\end{document}